\documentstyle[11pt,aaspp,psfig]{article}
\def\gs{\mathrel{\raise0.35ex\hbox{$\scriptstyle >$}\kern-0.6em \lower0.40ex\hbox{{$\scriptstyle \sim$}}}}
\def\ls{\mathrel{\raise0.35ex\hbox{$\scriptstyle <$}\kern-0.6em \lower0.40ex\hbox{{$\scriptstyle \sim$}}}}

\begin{document}
\small

\title{Optical spectral signatures of dusty starburst galaxies}
\author{
Bianca M.\ Poggianti$^{\!}$\altaffilmark{1,2,3}
\& Hong\ Wu$^{\!}$\altaffilmark{4,2}
}
\smallskip

\affil{\scriptsize 1) Osservatorio Astronomico di Padova, vicolo dell'Osservatorio 5, 35122 Padova, Italy}
\affil{\scriptsize 2) Institute of Astronomy, Madingley Rd, Cambridge CB3 OHA, UK}
\affil{\scriptsize 3) Royal Greenwich Observatory, Madingley Rd, Cambridge CB3 0EZ, UK}
\affil{\scriptsize 4) Beijing Astronomical Observatory, Chinese Academy of Sciences, Beijing 100080, China}

\begin{abstract}
We analyse the optical spectral properties of the complete sample of Very
Luminous Infrared Galaxies presented by Wu et al. (1998a,b) and 
we find a high fraction ($\sim$ 50 \%) of spectra showing both
a strong $\rm H\delta$ line in absorption and relatively modest [O{\sc ii}]
emission (e(a) spectra). The e(a) signature has been proposed 
as an efficient method to identify dusty starburst galaxies 
and we study the star formation activity and the
nature of these galaxies, as well as the effects of dust on their
observed properties. 
We examine their emission line characteristics, in particular 
their [O{\sc ii}]/$\rm H\alpha$ ratio, 
and we find this to be greatly affected by reddening.
A search for AGN spectral signatures reveals that the e(a)'s 
are typically HII/LINER galaxies.
We compare the star formation rates derived from the FIR luminosities
with the estimates based on the $\rm H\alpha$ line and find that the
values obtained from the optical
emission lines are a factor of 10-70 ($\rm H\alpha$)
and 20-140 ([O{\sc ii}]) lower than the FIR estimates (50-300 $M_{\odot} \, 
{yr}^{-1}$).
We then study the morphological properties of the e(a) galaxies, looking for 
a near companion or signs of a merger/interaction. 
In order to explore the evolution of the e(a) population,
we present an overview of the available observations of e(a)'s in 
different environments both at low and high redshift.
Finally, we discuss the role of dust in determining the e(a) spectral 
properties and we propose a scenario of selective obscuration in which
the extinction decreases with the stellar age.

\end{abstract}

\keywords{galaxies: clusters: general --- galaxies: evolution --- 
infrared: galaxies}

\sluginfo
\newpage

\section{Introduction}

The remarkable achievements of a number of recent optical observations of 
distant galaxies have delineated a picture of the star formation history of 
the Universe at redshift $0<z<5$ (see reviews from Ellis 1998a, 1998b and 
references therein; Steidel et al. 1999). At the same time, this 
progress has accentuated the necessity of evaluating the role played by dust
in obscuring star--forming systems (Meurer et al. 1997, Pettini et al. 1998,
Dickinson 1998, Steidel et al. 1998, Glazebrook et al. 1999)
and has strongly reminded the astronomical community
that relying on optical data alone is likely to offer a 
partial view of the evolutionary history of galaxies,
certainly quantitatively incomplete and possibly qualitatively incorrect.

A major remained challenge is thus to assess the amount of star formation
activity hidden by dust in the objects detected and uncover eventual 
entire galactic populations which might have been missed by the optical census.
Deep surveys in the submillimeter regime with SCUBA (Smail et al. 1997, 
1998a, Hughes et al. 1998, Barger et al. 1998,
Lilly et al. 1999, see Smail et al. 1998b for a review) 
and in the Mid/Far--Infrared with ISO (Flores et al. 1998a,b,
Elbaz et al. 1998, Altieri et al. 1999, Aussel et al. 1999, Puget et al.
1999, Clements et al. 1999, Metcalfe et al. 1999, Oliver et al. 1999) 
are starting to unveil high/intermediate 
redshift populations of galaxies whose star formation
is hidden/underestimated at optical wavelengths. 
A full exploitation of this observational break-through
at long wavelengths relies on the comparison with optical and near-infrared
data for the identification of the counterparts of these sources, for the 
determination of their redshift and for investigating the nature
of the source of the submillimeter/IR emission.

A key question is whether the distant, dusty star--forming galaxies are
recognizable from optical/near-IR data alone. 
Heavily reddened sources cannot be unambiguously
identified by optical and near-IR colors: photometry of even the most 
extreme cases ($R-K > 5-6$, EROs, Extremely Red Objects)
cannot establish whether these objects 
are heavily reddened starbursts at high-$z$ (Graham \& Dey 1996, 
Cimatti et al. 1998, Dey et al. 1999)
or high redshift analogs of local elliptical galaxies (Spinrad et al. 1997,
Cohen et al. 1999, Treu et al. 1999).
Thus finding an optical \it spectroscopic \rm signature able to pick out
highly extincted galaxies would be extremely valuable, especially given 
the large number of spectra accessible from recent and near-future 
spectroscopic surveys.

First evidence for the existence of such a signature has emerged 
from distant cluster studies.
Dressler et al. (1999, D99) presented a spectroscopic catalog
of cluster and field galaxies at $z \sim 0.4-0.5$
and identified a class of spectra with strong Balmer lines in absorption
and [O{\sc ii}] in emission (hereafter ``e(a)'' galaxies, 
EW($\rm H\delta) > 4$ \AA $\,$ and EW(O{\sc ii}$) > 
5$ \AA), which was found to be abundant at intermediate redshifts
both in the cluster and in the field environment
(about 10 \% of the entire galactic population in D99, see also
Poggianti et al. 1999, P99).
These authors also found a numerous \it cluster \rm population of galaxies
with strong Balmer lines in absorption 
and no detectable emission--lines (``k+a/a+k'' class, 
EW($\rm H\delta)_0 > 3$ \AA, traditionally named ``E+A'' galaxies),
which are universally interpreted as post--starburst/post--starforming
galaxies in which the star formation ended between
a few Myr and 1.5 Gyr before the time of observation (P99).

On the basis of dust--free models,
the e(a) spectra appear associated with \it post--starburst \rm
galaxies with some residual star formation (P99).
However, a search for similar spectra in the local Universe (P99)
revealed that they are frequent among merging/strongly interacting
galaxies (40 \%  in the sample of Liu \& Kennicutt 1995a,b, LK95a,b),
while they are scarce in normal field galaxy
samples (8--7 \% in the Las Campanas Redshift Survey and in
Kennicutt 1992a,b, K92a,b, see \S5.1 for details).
All of the e(a) galaxies in LK95 are strong FIR emitters and are
known to be highly extincted starbursts.

From the comparison of the distant e(a) galaxies with their low-$z$
counterparts and the similarity of their spectral properties (in particular, 
their low EW(O{\sc ii})/EW($\rm H\alpha$) ratios),
P99 concluded that the e(a) galaxies are starbursts which contain
a substantial amount of dust and suggested that they might be the progenitors
of the abundant cluster population of post--starburst galaxies (k+a/a+k).
They also tentatively explained the e(a) peculiar spectral properties
as the result of ``selective'' dust extinction, which affects the youngest,
most massive stars in HII regions much more than the older 
stellar populations responsible for the continuum flux.

It is clear that if the e(a) signature is proved to be unequivocally 
associated with 
highly extincted starbursts, then the identification of a strong $\rm H\delta$
line in emission--line spectra is a very powerful method to detect
dusty--starforming galaxies up to high redshift.
Since in this case the SFR estimates based on optical data are unreliable,
it is important to establish whether the detection of e(a) spectra,
besides revealing the \it presence \rm and \it incidence \rm
of a starburst population,
also allows to \it quantify \rm the amount of star formation
hidden by the dust and whether it provides definite 
informations about the properties and nature of these galaxies.
These final issues could not be investigated on the basis of the D99 distant
cluster sample, for which independent means of deriving the SFRs 
(e.g. FIR data) are not available. Hence a number of questions have 
remained open:

a) Are \it all \rm the e(a) spectra associated with dusty starbursts?
i.e. What is the range of FIR luminosities of the e(a) galaxies?

b) What fraction of the dusty starburst population display en e(a) spectrum?
i.e. What is the proportion of e(a) galaxies as a function of the FIR 
luminosity?

c) What are their typical star formation rates? What percentage is hidden 
by dust?

d) Which star formation and dust properties/geometries are responsible
for the e(a) spectral characteristics?

e) Are e(a) spectra always connected with merging galaxies, or
are there other mechanisms that can produce them?

f) What is the frequency of e(a) galaxies as a function of redshift and 
environment?

The goal of this paper is to investigate some of these issues
by analysing the sample of Very Luminous Infrared Galaxies
(VLIRGs, log($L_{IR}/L_{\odot}) \ge 11.5$) 
presented in Wu et al. (1998a,b W98a,b).
This sample has a number of advantages for the purpose of our study:
it is the largest, complete sample of VLIRGs obtained to date,
comprising a number of objects adequate for a statistical study. 
The wealth of information available for these galaxies (emission--line 
luminosities over a wide spectral range, FIR luminosities, 
merger stage/morphologies, reddening estimates) makes it an excellent
dataset for testing the connection between e(a) spectra and dusty
starbursts and for inspecting the properties of galaxies characterized by
this type of spectra. The ultimate purpose of this analysis 
is to investigate the nature of the e(a) galaxies in general and hopefully 
shed some light on the properties of the distant e(a)'s found in deep
spectroscopic surveys.

The plan of the paper is as follows:
in the next section we briefly present the spectroscopic catalog
and we describe the line measurements and 
the spectral classification (\S2.1).
In \S3 we show the results of our spectral analysis and we examine
the distributions of ``activity'' (AGN) type of our spectral classes. 
In \S4 we compare the spectral characteristics with other
galactic properties (FIR luminosities, reddening estimates, 
morphologies)
in order to establish the typical dust obscuration, star formation rates
and merger stage of the e(a) population.
In \S5 we review the incidence of e(a)'s in other spectroscopic surveys
both at low and high redshift and in \S6
we examine the possible physical scenarios accounting for the observed
spectral properties (\S6.1). We then summarize our main conclusions in \S7.
We assume $H_0 = 50 \, \rm km \, s^{-1} \, Mpc^{-1}$ and $q_0 = 0.5$.

\section{The spectroscopic catalog}

In this paper we will make use of the spectroscopic catalog of very 
luminous infrared galaxies obtained by W98a,b. W98a observed a complete sample
(73 galaxies, hereafter the ``VLIRG sample'') selected 
from the 2 Jy redshift survey (Strauss et al. 1992) 
with log($L_{IR}/L_{\odot}) \ge 11.5$ and sufficiently optically bright
to obtain good S/N spectra.
The 2 Jy survey includes all the objects from the IRAS Point Source catalog
with $f_{60} > 1.936$ Jy, $f^{2}_{60} > f_{12}f_{25}$ 
(a color criterion distinguishing galaxies from Galactic objects)
and Galactic latitude $|b| > 5^o$. 

In addition, spectra for 40 companions of VLIRG galaxies and 
for 10 galaxies with slightly lower FIR luminosity/fainter optical magnitudes
were obtained (W98a). In the following we will refer to the 
40 companion galaxies as the ``companion sample''
and to the total sample (VLIRG+companion) as the ``total W98 sample''. 
Most of these sources lie in the redshift range $0.02 < z < 0.05$
and have log($L_{IR}/L_{\odot})$ between 11.5 and 12
(see W98a for the redshift and infrared luminosity distributions).
For the purposes of this paper we have included into the VLIRG
sample 9 of the additional 10 galaxies with fainter/unknown optical magnitudes
(6 objects) or log($L_{IR}/L_{\odot}) \ge 11.465$ (3 objects), 
excluding only 1 galaxy
significantly fainter in the FIR (09517+6954, log($L_{IR}/L_{\odot})=10.833$). 
Three companion and one VLIRG spectra are too poor to determine
their redshift/spectral class and have been disregarded in the 
analysis.

The spectra were obtained with the 2.16 m telescope at the Beijing 
Astronomical Observatory and cover the spectral range either
4400--7100 \AA $\,$ or 3500--8100 \AA $\,$
at a resolution of 11.2 and 9.3 \AA $\,$ respectively.
\footnote{A full description of the data reduction
procedure and of the spectra obtained can be found in W98a.}
The apertures were varied according to the redshift
and correspond to the central 2 kpc for galaxies with
$z<0.034$, while for the most distant objects (about 40 \% of the sample) 
a larger area is covered.

In addition to the FIR luminosities
\footnote{The FIR luminosities given from W98a are computed from
the 60 $\micron$ and 100 $\micron$ fluxes ($F_{IR}=1.75[2.55S_{60}
+1.01S_{100}] 10^{-14} \, W \, m^{-2}$).
\it These are approximately total FIR luminosities in the range 
1--1000 $\micron$. \rm Following the definition of ``total far--infrared
emission'' given by Helou et al.(1988) 
and adopting a standard correction factor 1.5 to convert the 42.5--122.5 
$\micron$ emission into a 1--1000 $\micron$
flux, the factors (1.75;2.55;1.01) in the formula above
would be replaced by (1.89;2.58;1.00).}, W98a,b list
the relative fluxes of a number of emission--lines, the observed $\rm H 
\alpha$ fluxes, the equivalent widths (EWs) of $\rm H \alpha$ and  
the intrinsic reddening estimates.
We stress that, since many galaxies in the W98 sample are in pairs, 
multi--nuclei
systems or groups, in some cases the task of identifying the component
emitting the FIR luminosity is problematic. For most galaxies the
identification was done using the IRAS position and error ellipse,
while for those sources with more than one companion in the error
ellipse, the optical counterpart was found from the 
infrared colors or order of activity (Seyfert1, Seyfert2, LINER,
HII) (W98b).
For four of the sources (08507+3520, 13373+0105, 13458+1540, 13496+0221)
it was impossible to determine which companion accounts for
the FIR emission and this has been assigned to each one of the components.
For at least some of the sources, the FIR emission probably comes from two
or even more companions, hence the discrimination
between a VLIRG and a companion galaxy can be somewhat arbitrary 
and the companion sample should not be regarded in any way as a
``low-FIR luminosity'' sample.

\subsection{Spectral measurements and classification}

In order to study the star formation and dust properties of these 
galaxies it is useful to classify their spectra into a number of classes.
We adopt here a spectral classification scheme similar to the one presented
in D99. This scheme (Table 1) is based on two lines, 
[O{\sc ii}]$\lambda$3727 in emission 
and $\rm H\delta$($\lambda=4101$ \AA) in absorption, which are
good indicators of (respectively) current and recent star formation
in distant galaxy spectra.

We measured the strength of these two lines and of the other main
spectral features with two methods: 
using the package SPLOT within the IRAF environment (comparing the results
found with a Gaussian fit and interactively choosing the line endpoints)
and using the same purpose--written software
as in D99. The two methods gave consistent results, and 
in the case of the $\rm H\alpha + NII$ blend our 
measurements are in good agreement with the values listed in W98a,b.
Since we wish to compare our results with those obtained by D99,
in the following we will use the values 
derived with their method. Repeated measuments
show that the typical uncertainty in the EWs is $\sim 20 \%$
for the weakest lines and a few per cent for the strongest ones.

\begin{table*}
{\scriptsize
\begin{center}
\centerline{\sc Table 1}
\vspace{0.1cm}
\centerline{\sc Spectral Classification Scheme}
\vspace{0.3cm}
\begin{tabular}{lcccl}
\hline\hline
\noalign{\smallskip}
 {Class} & {EW([O{\sc ii}]\,3727)} & {EW(H$\delta$)} &  Comments \cr
 & (\AA) & (\AA) & &  \cr
\hline
\noalign{\smallskip}
k & absent & $< 3$ & passive \cr
k+a/a+k & absent & $\geq 3$ & strong Balmer absorption without emission \cr
e(c) & yes,$< 40$ & $< 4$ & moderate Balmer absorption plus emission, spiral-like \cr
e(a) & yes & $\geq 4$ & strong Balmer absorption plus emission \cr
e(b) & $\geq 40$ & any & very strong [O{\sc ii}] \cr
sey1 & any & any & Seyfert1 from broad hydrogen lines in emission\cr
\noalign{\smallskip}
e &  yes & ? & at least one emission line, $\rm H\delta$ unmeasurable \cr
\noalign{\smallskip}
\noalign{\hrule}
\noalign{\smallskip}
\end{tabular}
\end{center}
}
\vspace*{-0.8cm}
\end{table*}

\setcounter{table}{1}

Examples of the main emission--line classes 
classified according to the $H\delta$ and [O{\sc ii}]
strength are shown in Fig.~1. 
The $\rm H\delta$ line is prominent in absorption
in the e(a) spectrum (top panel, 6 \AA), while it is
absent/weak in emission in the other two spectra.

\hbox{~}
\vspace{-0.5in}
\centerline{\psfig{file=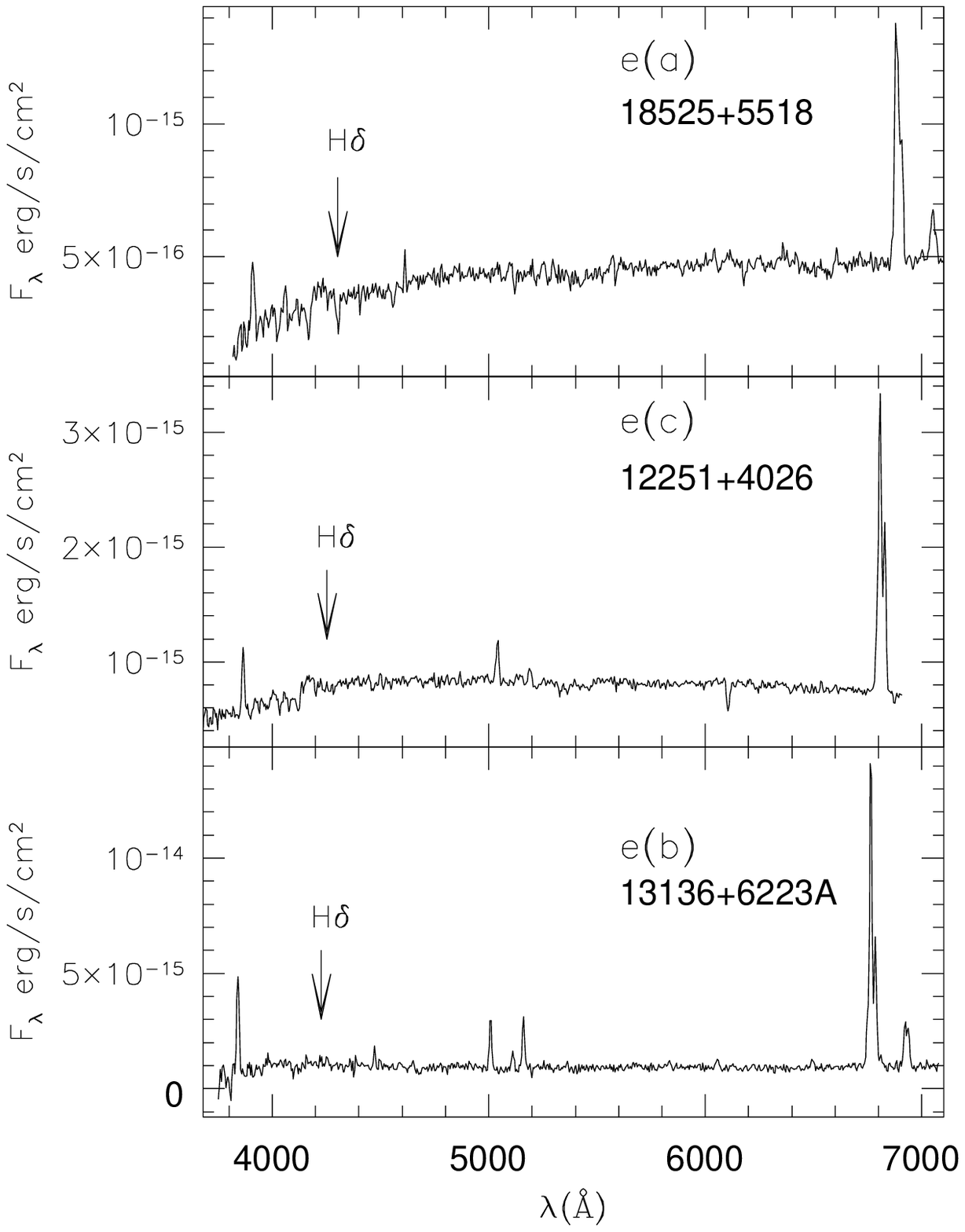,angle=0,width=4.0in}}
\vspace{-0.5in}

\noindent{\scriptsize
\addtolength{\baselineskip}{-3pt} 
\hspace*{0.1cm} Fig.~1.\ Representative spectra of the e(a), e(c) and e(b)
classes. A limited spectral range including both the 
[O{\sc ii}] and the $\rm H\alpha$ lines is displayed.
The position of the $\rm H\delta$ line is shown in each 
panel with an arrow. The complete spectroscopic catalog is presented in
W98a.

\addtolength{\baselineskip}{3pt}
}

A detailed modeling of each spectral class can be found 
in P99. Here we briefly summarize their conclusions
and proposed interpretation of each class:

\it k type: \rm passive, elliptical--like spectrum of a galaxy with 
neither on--going nor recent star formation.

\it k+a/a+k type: \rm spectrum of a post--starburst/post--starforming galaxy
with no current star formation which
was forming stars at a vigorous rate in the recent past (last 1.5 Gyr).

\it e(c) type: \rm typical spectrum of
a spiral at low redshift which has been forming stars
in a continuous fashion (without bursts) for the last 2 Gyr or longer.

\it e(a) type: \rm spectrum of a dust--enshrouded starburst galaxy.

\it e(b) type: \rm spectrum with very strong emission lines of a galaxy
which is undergoing strong star formation.

\it sey1 type: \rm Seyfert1 spectrum with broad emission lines.

\it e type: \rm spectrum of a galaxy with current star formation. Since the
$\rm H\delta$ line is unmeasurable (in most cases because it falls out
of the observed spectral range) we cannot classify it into one
of the emission line classes described above.

\section{Results}

\begin{table*}
{\scriptsize
\begin{center}
\centerline{\sc Table 2}
\vspace{0.1cm}
\centerline{\sc VLIRG galaxies}
\vspace{0.1cm}
\begin{tabular}{llclrrrrrrr}
\hline\hline
\noalign{\smallskip}
 IRAS name & spec. & $L_{IR}/L_{\odot}$$^{a}$ & AGN & EW [O{\sc ii}] & EW($\rm H\beta$) &
EW($\rm H\alpha + NII$) & EW($\rm H\delta$) & EW [O{\sc iii}]$^{c}$ & SFR$_{FIR}$$^{d}$ & SFR$_{\rm H\alpha}$$^{d}$ \cr
 & class & (log) & type$^{b}$ & (\AA) & (\AA) & (\AA) & (\AA) & (\AA) & $M_{\odot}/yr$ & $M_{\odot}/yr$ \cr
(1) & (2) & (3) & (4) & (5) & (6) & (7) & (8) & (9) & (10) & (11) \cr
\hline
\noalign{\smallskip}
00189+3748A  &  e(a)    &  11.572 & LH  & 14.3  &  7.2  &  86.5  &  4.6  & 1.4 & 64.2  &   2.17 \cr
00267+3016A  &  e(a)    &  11.966 & L   & 10.9  &  2.4  &  75.6  &  4.1  & 4.1 & 159.0  &   5.51 \cr
00509+1225   &  sey1    &  11.772 & S1  &  3.1  & 19.9  & 143.4  & -2.6  & 14.6& 101.7  &  42.89 \cr
01173+1405   &  e(b)    &  11.868 & LH  & 42.5  & 11.8  & 145.6  &  0.0  & 10.6& 126.9  &   1.12 \cr
01324+2138   &  e(a)    &  11.629 & LH  & 25.8  &  4.7  &  75.8  &  7.9  & 15.6: & 73.2  &   2.27 \cr
01484+2220   &  e       &  11.851 & H   &  out  &  5.3  &  59.0  &  out  & 1.2 & 122.0  &   1.61 \cr
01572+0009   &  sey1    &  12.665 & S1  &  2.5  & 20.2  & 177.3  &  fil  & 32.3 & 795.3  & 163.36 \cr
02071+3857   &  e(a)    &  11.546 & H   &  out  &  fil  &  34.1  & 13.3  & 0.8 &  60.5  &   1.08 \cr
02203+3158A  &  e(a)    &  11.837 & LH  &  6.2  &  2.4  &  46.2  &  6.5  & 4.5 & 118.2  &   1.94 \cr
02222+2159   &  e       &  11.652 & H   &    ?  &  0.0  &  35.3  &  0.0  & 2.1 & 77.2  &   0.98 \cr
02248+2621   &  e(c)    &  11.519 & H   &  9.0  &  9.5  & 160.7  &    ?  & 5.0 & 56.8  &   6.94 \cr
02435+1253   &  e(a)    &  11.501 & LH  & 25.6: &  0.0  &  43.4  &  9.7  & 5.1 & 54.5  &   0.24 \cr
02512+1446A  &  e       &  11.780 & H   & 12.6  & 10.9  & 141.2  &    ?  & 5.3: & 103.6  &   1.73 \cr
03117+4151A  &  e(b)    &  11.562 & S2  & 66.7  & 17.4  & 207.1  &  0.0  & 93.8 & 62.7  &   5.25 \cr
05084+7936   &  e(a)    &  12.170 & H   & 35.6  &  8.0  & 101.8  &  7.8  & ? & 254.4  &   6.62 \cr
05414+5840   &  e       &  11.505 & S?  &  out  & -7.7  &  17.7  &  0.0  & ? &  55.0  &   0.06 \cr
06538+4628   &  e       &  11.490 & H   &  out  & 12.0  & 128.2  &    ?  & 6.2 & 53.2  &   4.61 \cr
07062+2041B  &  e(a)    &  11.559 & H   & 18.2  &  5.9  &  64.5  &  4.9  & 4.1 & 62.3  &   1.47 \cr
07063+2043A  &  e(a):   &  11.570 & H   &  out  &  4.3  &  70.1  &  8.2: & 2.5 & 63.9  &   1.55 \cr
07256+3355A  &  e(a)    &  11.467 & H   &  6.9  &  3.4  &  66.6  &  5.3  & 1.2: & 50.4  &   1.14 \cr
08354+2555   & e(a)$^+$ &  11.781 & S?  &  0.0  & -5.3  &  23.0  &  7.2  & --- & 103.9  &   0.32 \cr
08507+3520A  &  e(c)    &  11.811 & S?  &  0.0  &  4.3  &  22.8  &  0.0  & 3.1: & 111.3  &   1.03 \cr
08507+3520B  &  e(b,a): &  11.811 & S?  & 73.0  &  5.5  &  44.7  &  9.2: & 9.7: & 111.3  &   1.31 \cr
08507+3520C  &  e       &  11.811 & S?  &$<19.1$&  7.7  &  55.9  &    ?  & 10.1 & 111.3  &   1.00 \cr
09047+1838   &  e(a)    &  11.490 & H   & 11.6  &  4.7  &  59.7  &  4.2  & 1.5 & 53.2  &   1.22 \cr
09126+4432A  &  e(a):   &  11.913 & LH  & 10.3  &  0.0  &  29.2  &  1.5: & 1.5 & 140.8  &   1.03 \cr
09168+3308   &  e(a):   &  11.725 & H   & 18.0  &  8.8  &  84.2  &    ?  & 4.0 & 91.3  &   3.41 \cr
09320+6134   &  e(a)    &  12.220 & L   & 18.9  &  fil  &  63.5  &  8.6  & 4.6 & 285.4  &   1.79 \cr
09333+4841A  &  e(a)    &  11.523 & H   & 16.1  &  3.0  &  60.9  &  5.7  & 3.6 & 57.3  &   1.26 \cr
10203+5235   &  e(c)    &  11.620 & L   & 25.6  &  5.9  &  71.7  &  3.6  & 8.2 & 71.7  &   1.94 \cr
10311+3507   &  e       &  12.096 & H   &  out  & 13.5  & 113.5  &  out  & 2.7 & 214.5  &  15.25 \cr
11231+1456A  &  e(c)    &  11.809 & LH  & 14.3  &  5.9  &  84.9  &  2.3  & 1.1: & 110.8  &   1.90 \cr
11254+1126   &  e       &  11.800 & LH  &  out  &  5.3  &  89.9  &  out  & 8.6 & 108.5  &   4.48 \cr
11257+5850B  &  e       &  12.040 & LH  &  out  & 30.6  & 219.5  &  out  & 29.7 & 188.6  &   0.67 \cr
11543+0124   &  e(a)    &  11.716 & LH  & 13.8: &  0.0  &  40.5  &  6.1  & 2.0 & 89.4  &   1.50 \cr
12112+0305A  &  e(c)    &  12.531 & LH  & 35.2  &  5.3  &  88.2  &  fil  & 13.3 & 584.2  &   3.21 \cr
12120+6838A  &  e       &  12.029 & LH  &  out  &  4.3  &  53.6  &  out  & 4.1 & 183.9  &   2.33 \cr
12251+4026   &  e(c)    &  11.660 & H   & 13.3  &  6.5  &  91.5  &  fil/emi  & 3.7 & 78.6  &   2.96 \cr
12265+0219   &  sey1    &  12.663 & S1  &  0.0  & 40.3  & 203.8  &  emi  & 6.3: & 791.6  &2316.27 \cr
12323+1549B  &  e(c)    &  11.766 & S1  &  6.7  & 15.9  & 104.9  &  0.0  & 8.8 & 100.4  &   8.23 \cr
12540+5708   &  sey1    &  12.635 & S1  &  out  &    ?  & 122.8  &  out  & 6.2: & 742.2  & 188.40 \cr
12592+0436   &  e(a)    &  11.787 & L   &  4.6  & -6.9  &  28.8  &  4.4  & ? & 105.3  &   0.40 \cr
13136+6223A  &  e(b)    &  11.937 & H   &102.2  & 34.8  & 203.5  &  emi  & 27.8: & 148.8  &   8.86 \cr
13183+3423   &  e       &  11.863 & LH  &  out  &  1.0  &  88.3  &  out  & 4.5 & 125.5  &   0.97 \cr
13299+1121   &  e       &  11.516 & H   &  out  & 28.6  & 235.0  &  out  & 17.1 &  56.4  &   9.14 \cr
13362+4831B  &  e       &  11.706 & S2  &  out  & 15.2  & 160.7  &  out  & 87.7 & 87.4  &   2.05 \cr
13373+0105A  &  e       &  11.701 & LH  &  out  &    ?  &  17.9  &  out  & --- & 86.4  &   0.22 \cr
13373+0105B  &  e       &  11.701 & H   &  out  & 19.4  & 144.3  &  out  & 14.4 & 86.4  &   1.36 \cr
13428+5608   &  e       &  12.392 & S2  &  out  &  9.9  & 205.4  &  out  & 60.2 & 424.2  &   5.95 \cr
13458+1540A  &  e(a)    &  11.821 & H   & 13.3  &  5.8  &  98.4  &  4.9  & 2.9 & 113.9  &   5.02 \cr
13458+1540B  &  e(c)    &  11.821 & H   & 13.7  &  4.9  &  68.6  &  fil  & 4.7 & 113.9  &   2.53 \cr
13496+0221A  &  e(a)    &  11.752 & H   &  8.9  &  0.0  &  24.2  &  4.5  & --- & 97.2  &   0.70 \cr
13496+0221B  &  e       &  11.752 & H   & 25.5  &  3.2  &  30.3  &  4.6: & --- & 97.2  &   0.92 \cr
13536+1836   &  e       &  11.611 & S2  &  out  & 37.5  & 457.3  &  out  & 312.2 & 70.2  &  25.89 \cr
\noalign{\smallskip}
\noalign{\hrule}
\end{tabular}
\end{center}
}
\vspace*{-0.8cm}
\end{table*}

\begin{table*}
{\scriptsize
\begin{center}
\centerline{\sc Table 2 -- continued}
\vspace{0.1cm}
\centerline{\sc VLIRG galaxies}
\vspace{0.1cm}
\begin{tabular}{llclrrrrrrr}
\hline\hline
\noalign{\smallskip}
 IRAS name & spec. & $L_{IR}/L_{\odot}$$^{a}$ & AGN & EW 
[O{\sc ii}] & EW($\rm H\beta$) & EW($\rm H\alpha + NII$) & 
EW($\rm H\delta$) & EW [O{\sc iii}]$^{c}$ & SFR$_{FIR}$$^{d}$ & SFR$_{\rm H\alpha}$$^{d}$ \cr
 & class & (log) & type$^{b}$ & (\AA) & (\AA) & (\AA) & (\AA) & (\AA) &  
$M_{\odot}/yr$ & $M_{\odot}/yr$ \cr
(1) & (2) & (3) & (4) & (5) & (6) & (7) & (8) & (9) & (10) & (11)  \cr
\hline
\noalign{\smallskip}
14151+2705A  &  e(a)    &  11.565 & LH  &  3.4: &$<-7.7$&  24.3  &  5.4  & 1.0: & 63.2  &   2.04 \cr
14178+4927   &  e(a):   &  11.541 & LH  &  0.0  &  4.3  &  72.2  &  3.9  & 1.4 & 59.8  &   1.10 \cr
14547+2448A  &  e       &  11.897 & L   &  out  &  2.5  &  36.8  &  out  & 6.2 & 135.7  &   0.57 \cr
14568+4504   &  e       &  11.501 & LH  &  out  &  4.4  &  57.7  &  out  & 8.1 & 54.5  &   2.79 \cr
15107+0724   &  e(c)    &  11.525 & H   &  out  &  0.0  &  30.7  &    ?  & --- & 57.6  &   0.08 \cr
15163+4255   &  e       &  12.072 & H   &  out  & 17.4  & 235.4  &  out  & 15.2 & 203.0  &   8.27 \cr
15327+2340   &  e       &  12.464 & S?  &  out  &    ?  &  44.1  &  out  & 13.3: & 500.6  &   0.06 \cr
15425+4114A  &  e       &  11.515 & H   &  out  &  5.8  &  79.9  &  out  & 7.3 & 56.3  &   2.28 \cr
15426+4116B  &  e       &  11.546 & LH  &  out  &  3.2  &  10.4  &  out  & --- & 60.5  &   0.53 \cr
16104+5235   &  e       &  11.687 & H   &  out  & 34.7  & 328.0  &  out  & 14.4 & 83.7  &  12.90 \cr
16180+3753   &  e       &  11.592 & H   &  out  &  7.4  &  44.4  &  out  & ? & 67.2  &   1.74 \cr
16284+0411   &  e       &  11.582 & LH  &  out  &  5.0  &  87.2  &  out  & --- & 65.7  &   1.13 \cr
16504+0228   &  e       &  12.028 & L   &  out  & 17.5  & 269.3  &  out  & 19.4 & 183.5  &   1.55 \cr
16577+5900A  &  e       &  11.582 & LH  &  out  &  3.9  &  31.1  &  out  & --- & 65.7  &   0.26 \cr
16589+0521   &  e(c)    &  11.637 & H   & 21.3  &  7.7  &  89.7  &  0.0  & 3.6 & 74.6  &   2.99 \cr
17366+8646   &  e(a):   &  11.544 & H   & 10.2  &  1.7  &  56.9  &  4.0  & 2.0 & 60.2  &   1.08 \cr
17392+3845   &  e       &  11.554 & H   &  out  &  4.8  &  48.7  &  out  & ? & 61.6  &   2.98 \cr
17501+6825A  &  e       &  11.829 & H   &  out  &  8.9  &  82.3  &  out  & 5.2 & 116.0  &   4.64 \cr
18525+5518   &  e(a)    &  11.683 & H   & 19.0  &  3.3  &  53.5  &  6.0  & 4.3: & 82.9  &   2.16 \cr
18595+5048   &  e(c)    &  11.501 & LH  & 17.2  &  1.6  &  44.2  &  0.0  & 1.9 & 54.5  &   0.77 \cr
19120+7320A  &  e       &  11.624 & S2  & 12.1  &  4.7  &  87.9  &  2.0: & 21.5 & 72.4  &   1.33 \cr
20550+1656   &  e(b)    &  12.074 & H   & 75.6  & 38.7  & 305.2  & -5.4  & 83.4 & 204.0  &  19.44 \cr
22388+3359   &  e(c):   &  11.531 & LH  &  7.5  &  fil  &  39.4  &  3.4: & --- & 58.4  &   0.41 \cr
22501+2427   &  e(c)    &  11.723 & H   &  1.5  & 10.7  & 162.1  &  fil  & 4.2 & 90.9  &   6.73 \cr
23007+0836A  &  sey1    &  11.734 & S1  &  3.0  & 22.4  & 174.6  &  emi  & 27.1 & 93.2  &  10.03 \cr
23024+1916   &  e       &  11.573 & LH  &  out  &  5.2  &  64.8  &  out  & 5.6 & 64.3  &   0.64 \cr
23135+2516   &  e(a)    &  11.730 & LH  & 16.2  &  2.0  &  52.3  &  4.1  & 8.6 & 92.4  &   1.11 \cr
23254+0830A  &  e       &  11.568 & H   &  7.0  &    ?  &  35.0  &    ?  & --- & 63.6  &   2.55 \cr
23488+1949A  &  e       &  11.528 & H   &  4.5  &  4.0  &  60.4  &  1.5: & 1.1 & 58.0  &   1.56 \cr
23488+2018A  &  e(a)    &  11.609 & H   & 11.9  &  5.6  &  93.1  &  4.1  & 2.7: & 69.9  &   1.71 \cr
23532+2513A  &  e(a):   &  11.795 & H   &    ?  &  abs  &  31.5  &  5.5: & ? & 107.3  &   1.12 \cr
23594+3622   &  e(a):   &  11.586 & S2  & 26.5  &  abs  &  62.5  &  2.5: & 4.4 & 66.3  &   0.69 \cr
\noalign{\smallskip}
\noalign{\hrule}
\noalign{\smallskip}
\multispan9{~~~$^{a}$ From W98a. \hfil}\cr
\noalign{\vspace{0.05cm}}
\multispan9{~~~$^{b}$ From W98b, see \S3.1. \hfil}\cr
\noalign{\vspace{0.05cm}}
\multispan9{~~~$^{c}$ [O{\sc iii}]$\lambda$5007. \hfil}\cr
\noalign{\vspace{0.05cm}}
\multispan9{~~~$^{d}$ SFR estimated from the FIR and $\rm H\alpha$
luminosity, see \S4.2. \hfil}\cr
\noalign{\vspace{0.05cm}}
\multispan9{EWs are given in the rest frame. \hfil}\cr
\noalign{\vspace{0.05cm}}
\multispan9{A colon mark ``:'' indicates an uncertain value or class. \hfil}\cr
\noalign{\vspace{0.05cm}}
\multispan9{``e(a)$^+$'' = $\rm H\alpha$ emission, no detectable [O{\sc ii}]
and strong $\rm H\delta$ absorption. \hfil}\cr
\noalign{\vspace{0.05cm}}
\multispan9{``out''= line unmeasurable because out of the spectrum range. \hfil}\cr
\noalign{\vspace{0.05cm}}
\multispan9{``fil''= line filled by the emission component. \hfil}\cr
\noalign{\vspace{0.05cm}}
\multispan9{``emi''= line in emission but unmeasurable. \hfil}\cr
\noalign{\vspace{0.05cm}}
\multispan9{``?''= line unmeasurable. \hfil}\cr
\end{tabular}
\end{center}
}
\vspace*{-0.8cm}
\end{table*}
	             	    	   		      	     
\begin{table*}
{\scriptsize
\begin{center}
\centerline{\sc Table 3}
\vspace{0.1cm}
\centerline{\sc Companion galaxies}
\vspace{0.3cm}
\begin{tabular}{lllrrrrr}
\hline\hline
\noalign{\smallskip}
 IRAS name & spec. & AGN & EW [O{\sc ii}] & EW($\rm H\beta$) &
EW($\rm H\alpha + NII$) & EW($\rm H\delta$) & EW [O{\sc iii}]$^{c}$   \cr
 & class & type$^b$ & (\AA) & (\AA) & (\AA) & (\AA) & (\AA) \cr
(1) & (2) & (3) & (4) & (5) & (6) & (7) & (8) \cr
\hline
\noalign{\smallskip}
00189+3748B  & e(a):         &   LH  & 15.2  &  3.1  &  46.2  & 11.2  & 2.2 \cr
00267+3016B  & e(a)          &   LH  &  8.9  &  3.0  &  74.4  & $>7.3$ & 5.5:  \cr
02203+3158B  & e(a)          &   H   &  5.1  &  2.3  &  38.4  &  4.6 & --- \cr
07256+3355B  & e(c)          &   H   &  0.0  &  4.3  &  34.9  &  2.3 & --- \cr
07256+3355C  & k             &   O   &  0.0  &  0.0  &   0.0  &  0.0 & --- \cr
09126+4432B  & e(a):         &   H   &    ?  &  emi  &  40.8  &  9.8 &  ? \cr
09333+4841B  & e             &   H   & 32.5: &  2.4  &  26.2  &    ? & 5.7: \cr
11231+1456B  & e(c)          &   LH  &  5.8  &  4.1  &  50.4  &  0.0 & 1.5: \cr
11257+5850A  & e             &   H   &  out  & 36.6  & 217.4  &  out & 30.7 \cr
12112+0305C  & e(c)          &   LH  & 24.2  &  6.8  &  76.1  &  0.0 & 15.5 \cr
12120+6838C  & e(a):$^+$     &   LH  &  0.0  &  0.0  &  12.3  &  4.1:& --- \cr
12120+6838D  & e             &   H   & 21.3  &  5.9  &  49.3  &    ? & 4.1 \cr
12323+1549A  & e(c)          &   LH  &    ?  &  0.0  &   9.6  &  1.6 & --- \cr
13136+6223B  & e             &   LH  &    ?  &    ?  &  57.4  &    ? & 8.9: \cr
13362+4831A  & e             &   L   &  out  & 24.9  & 225.4  &  out & 41.3 \cr
13496+0221C  & e             &   H   & 24.8  &  7.9  &  30.5  &    ? & --- \cr
14151+2705B  & e(a):         &   H   & 21.6: &  0.0  &   5.1: &  9.0:& --- \cr
14547+2448B  & e(a):         &   H   & 19.1: &  8.0  &  49.7  &  5.3:& --- \cr
16577+5900B  & e             &   LH  &  out  &  2.1  &  35.0  &  out & 3.3 \cr
17501+6825B  & e(c)          &   LH  & 26.7  &  5.8  &  71.4  &  fil & 14.0 \cr
17501+6825C  & e(a)          &   H   & 14.6: &  2.6: &  32.3  &  5.4 & 3.0: \cr
17501+6825D  & e(c)          &   LH  & 11.1  &  4.9  &  43.2  &  3.0:& 1.4: \cr
19120+7320B  & e(c)          &   H   & 13.5  & 10.2  &  95.5  &  fil & 1.4: \cr
23007+0836B  & e(c)          &   S?  &  0.0  &  fil  &  40.8  &  0.0 & ? \cr
23254+0830B  & e(b)          &   S2  & 45.6  & 25.4  & 263.0  &  fil & 248.7: \cr
23254+0830C  & e             &   O   &  5.3  &    ?  &   0.0  &    ? & ? \cr
23254+0830D  & k             &   O   &  0.0  & -2.8  &   0.0  &  1.2 & ? \cr
23488+1949B  & e(c)          &   H   & 33.9  & 15.8  & 146.4  &  0.0 & 13.1 \cr
23488+1949C  & e(a):$^+$     &   LH  &  0.0  &  3.9  &  61.6  &    ? & --- \cr
23488+2018B  & e             &   H   &  out  &  9.1  &  35.9  &  out & 15.6 \cr
23532+2513B  & sey1          &   S1  &  0.0  & 31.4  & 120.8  &  emi & 8.6 \cr
23532+2513C  & e(a):         &   H   & 12.1  &  4.3  &  59.5  &  8.1:& 6.1 \cr
\noalign{\smallskip}
\noalign{\hrule}
\noalign{\smallskip}
\multispan7{Notes as in Table 2. \hfil}\cr
\noalign{\smallskip}
\multispan7{IR07256+3355C, IR12323+1549A and IR23254+0830C should be 
unrelated to the FIR \hfil}\cr
\noalign{\smallskip}
\multispan7{source (background/foreground galaxies). \hfil}\cr
\end{tabular}
\end{center}
}
\vspace*{-0.8cm}
\end{table*}

In this section we present the results of our spectral analysis and we assess
the incidence of each spectral class among Very Luminous Infrared Galaxies.
By comparing the strength and the correlations between the various
spectral features with those observed in optically selected samples, 
we aim at studying the stellar and dust properties of these galaxies and 
the eventual presence of an AGN component in their spectra.

The spectral classification and our measurements of the
rest frame equivalent widths are shown in Table 2 (VLIRG sample) and
Table 3 (companion sample). 
Lines seen in emission are given a positive equivalent width, 
with the exception of the $\rm H\delta$ line 
which is given a positive value when it is observed in 
absorption. 

In Table 4 we list the proportion of galaxies as a function of the
spectral class; they are 
normalized to the total number of spectra with a \it securely assigned spectral
type, \rm excluding the ``e'' spectra that are
40\% and 28\% of the VLIRG and companion sample respectively.
The errors quoted are found assuming Poissonian statistics.

Table 4 shows that
all the VLIRGs and the great majority of the companion galaxies
have emission lines and this confirms that -- at least qualitatively --
the FIR and the optical data agree on revealing the existence
of on-going star formation activity. 
Among the VLIRGs the incidence of e(a) spectra 
\footnote{Three galaxies with a strong $\rm H\delta$ line in absorption and 
no detectable [O{\sc ii}] line have been included into the e(a) class 
because they display a clear $\rm H\alpha$ line in emission ($\rm e(a)^+$).
We note that at higher redshift ($z \ge 0.4$) these spectra would be 
classified as k+a galaxies because the $\rm H\alpha$ line
is redshifted out of the optical range; hence the ``true'' e(a) fraction 
in distant surveys could be higher than it is observed (P99).}
is very high (56 \%). About 1/4 of the VLIRGs have an e(c) ``spiral--like''
spectrum, while only 1 out of 10
have very strong [O{\sc ii}] emission (e(b) class). Broad hydrogen lines
denoting Seyfert1 spectra are observed in 10 \% of the VLIRGs.

Similar spectral fractions are found among the companion galaxies 
but with a slightly lower e(a) proportion (43 \%), an additional passive 
component (k class, 9 \%) and a higher proportion of normal ``spiral--like'' 
spectra (e(c) class, 39 \%).
Notably no k+a/a+k spectrum is found in the whole W98 catalog; we will comment
this result later in \S6. 
It is interesting that there is a strong tendency for e(a) galaxies 
to have e(a) companions: from Table 2 we find 6 e(a)--e(a) associations,
3 e(c)--e(c) companions and only 1 ``mixed couple'' e(a)--e(c).

\begin{table*}
{\scriptsize
\begin{center}
\centerline{\sc Table 4}
\vspace{0.1cm}
\centerline{\sc Fraction of galaxies as a function of the spectral class}
\vspace{0.3cm}
\begin{tabular}{lcrcrc}
\hline
\noalign{\smallskip}
Class & VLIRG & $N_{VLIRG}$ & compan. & $N_{compan.}$ & total \\
\hline
\noalign{\smallskip}
e(a) & 0.56$\pm$0.10 & 29 & 0.43$\pm$0.14 & 10 & 0.52$\pm$0.08 \\
e(c) & 0.25$\pm$0.07 & 13 & 0.39$\pm$0.13 &  9 & 0.29$\pm$0.06 \\
e(b) & 0.10$\pm$0.04 &  5 & 0.04$\pm$0.04 &  1 & 0.08$\pm$0.03 \\
sey1 & 0.10$\pm$0.04 &  5 & 0.04$\pm$0.04 &  1 & 0.08$\pm$0.03 \\
k    & 0             &  0 & 0.09$\pm$0.06 &  2 & 0.03$\pm$0.02 \\
\noalign{\smallskip}
\noalign{\hrule}
\noalign{\smallskip}
\end{tabular}
\end{center}
}
\vspace*{-0.8cm}
\end{table*}

The equivalent widths of [O{\sc ii}] and $\rm H\delta$ of the W98 galaxies
are plotted in the top panel of Fig.~2, where
the e(c), e(b) and k spectra have been grouped into the ``non--e(a)''
class. The EWs of the merging galaxies (LK95) (Fig.~2, lower panel)
are similar to those of the Infrared Luminous galaxies,  but the merger sample
has a greater proportion of non--e(a) galaxies.
The normal galaxies (K92a) occupy a 
confined region of the diagram with EW($\rm H\delta) < 4$ \AA $\,$ and 
EW([O{\sc ii}])$< 20$ \AA.
The EW([O{\sc ii}])s of e(a) and ``normal'' galaxies 
tend to increase with EW($\rm H\delta$) (Fig.~2) and this suggests
that the production and/or visibility of the nebular emission is related 
to the abundance and/or visibility of the stellar populations
responsible for the strong $\rm H\delta$. 

\hbox{~}
\centerline{\psfig{file=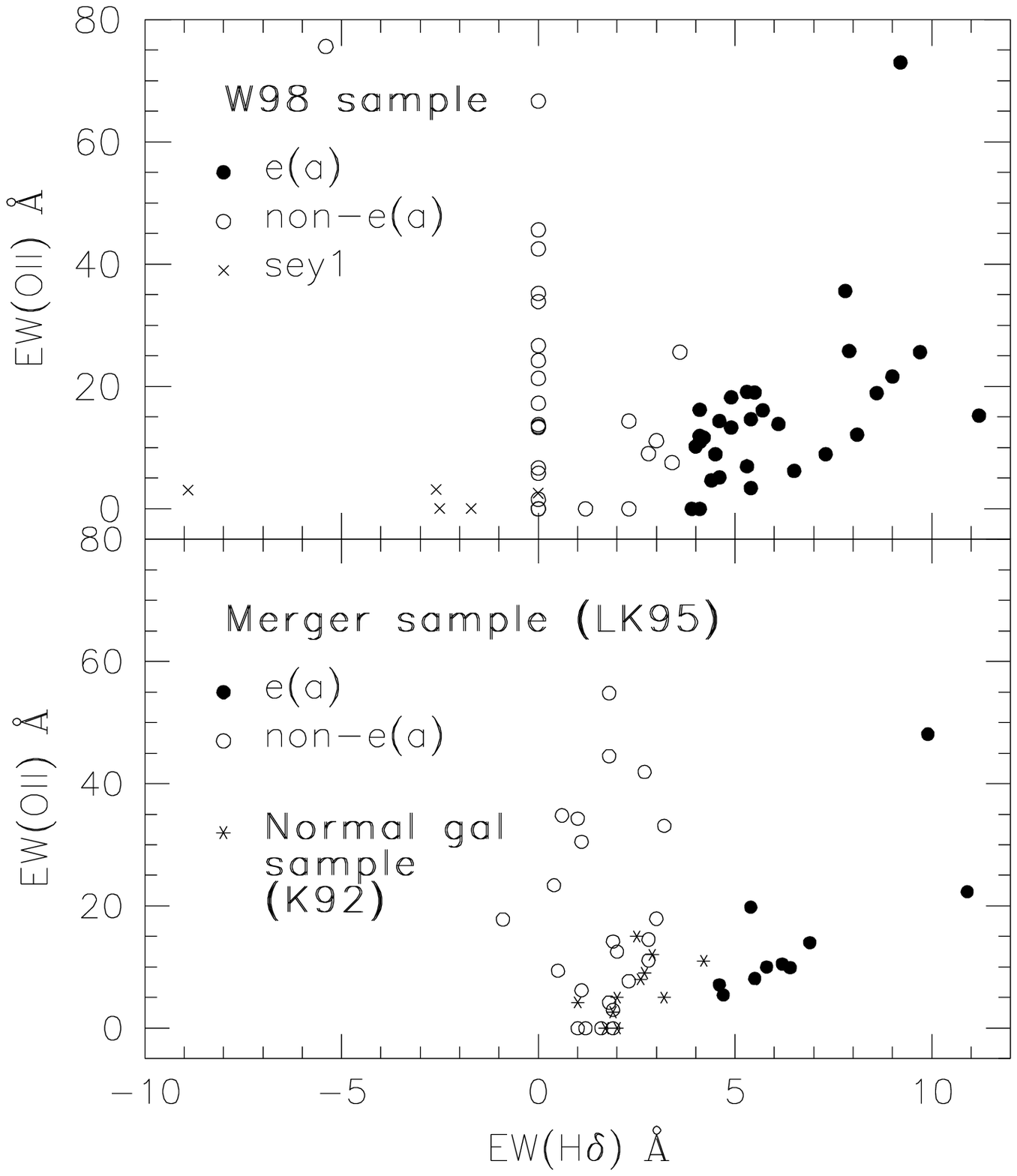,angle=0,width=4.0in}}
\vspace{-0.5in}

\noindent{\scriptsize
\addtolength{\baselineskip}{-3pt} 
\hspace*{0.1cm} Fig.~2.\ EW([O{\sc ii}]) versus EW($\rm H\delta$)
for the total W98 sample (top panel), the merging galaxies 
sample from LK95 (bottom panel; whole aperture/integrated (when available) or
nuclear spectra)
and the normal galaxies from K92 (types E to Sc with
no sign of a starburst or interaction/merger, bottom panel; spectra are
integrated).
The non--e(a) spectra (e(c),e(b),k) are represented by the empty dots.
Since the program that computes the line strength is not suitable to measure
weak or undetectable lines, in the top panel a value =0 have been assigned to 
the weakest $\rm H\delta$ equivalent widths, which in any case are smaller 
than 2 \AA. Note that most of the Seyfert1 spectra (crosses in the top panel) 
display the $\rm H\delta$ line in emission and have a very weak 
[O{\sc ii}] line.
\addtolength{\baselineskip}{3pt}
}

A notable property of the e(a) galaxies appears to be a
low EW(O{\sc ii})/EW($\rm H\alpha$+NII) ratio; this has been observed in
the e(a) spectra of distant cluster galaxies and of nearby mergers (P99).
We find the same characteristic also in the galaxies of the W98 sample:
in the EW-EW diagram (Fig.~3) the great majority of points 
lie on the right side of the straight line, which represents a rough fit
for normal field galaxies at low--redshift.
The median EW([O{\sc ii}])/EW($\rm H\alpha$+NII) ratios of the VLIRG
and the total W98 sample are 0.21$\pm$0.04 and 0.20$\pm$0.03 respectively.
These are a factor of 2 lower than the ratio found for optically--selected
nearby galaxies (0.42--0.47, K92, Tresse et al. 1999).
In the scenario proposed by P99, these low EW ratios 
are due to high dust extinction: the HII regions where the line 
emission is produced are deeply embedded in large amounts of dust, 
which affects the [O{\sc ii}] line more than $\rm H\alpha$.
Since the continuum underlying the lines is produced by older,
less extincted stars, the net result is a low EW ratio of the two lines. 
It is remarkable that in the Infrared Luminous sample
\it both the e(a) and the non--e(a) spectra have low ratios,
\rm while in the optically--selected sample of galaxies in distant clusters
the e(c) spectra are preferentially found along the fit for 
normal nearby galaxies (P99). This suggests that 
dust effects are not relevant in the e(c) spectra of the distant galaxies,
while strong extinction alters the emission line properties 
of all Infrared Luminous galaxies of any spectral class. 

Furthermore the median \it flux \rm ratios in W98 are 
F([O{\sc ii}])/F($\rm H\alpha$)=0.23$\pm$0.18 (Seyfert1's excluded). 
These are again a factor of $\sim 2$ lower than the median
flux ratio observed in nearby spirals (0.43$\pm$0.27, K92b).
\footnote{Note that the value NII/$\rm H\alpha$ can be 
neglected in this discussion because it depends only weakly on
EW($\rm H\alpha$) and it is approximately constant (K92, Tresse et al. 1999).}
The difference in \it equivalent width \rm ratio between the W98 sample
and normal nearby galaxies is hence entirely due to the difference
in the average \it flux \rm ratio of the two lines.
This is consistent with the suggestion that the main cause of 
the low EW([O{\sc ii}])/EW($\rm H\alpha$ + NII) is a differential
effect of dust extinction on the lines and the underlying continuum.

\hbox{~}
\vspace{-1.5in}
\centerline{\psfig{file=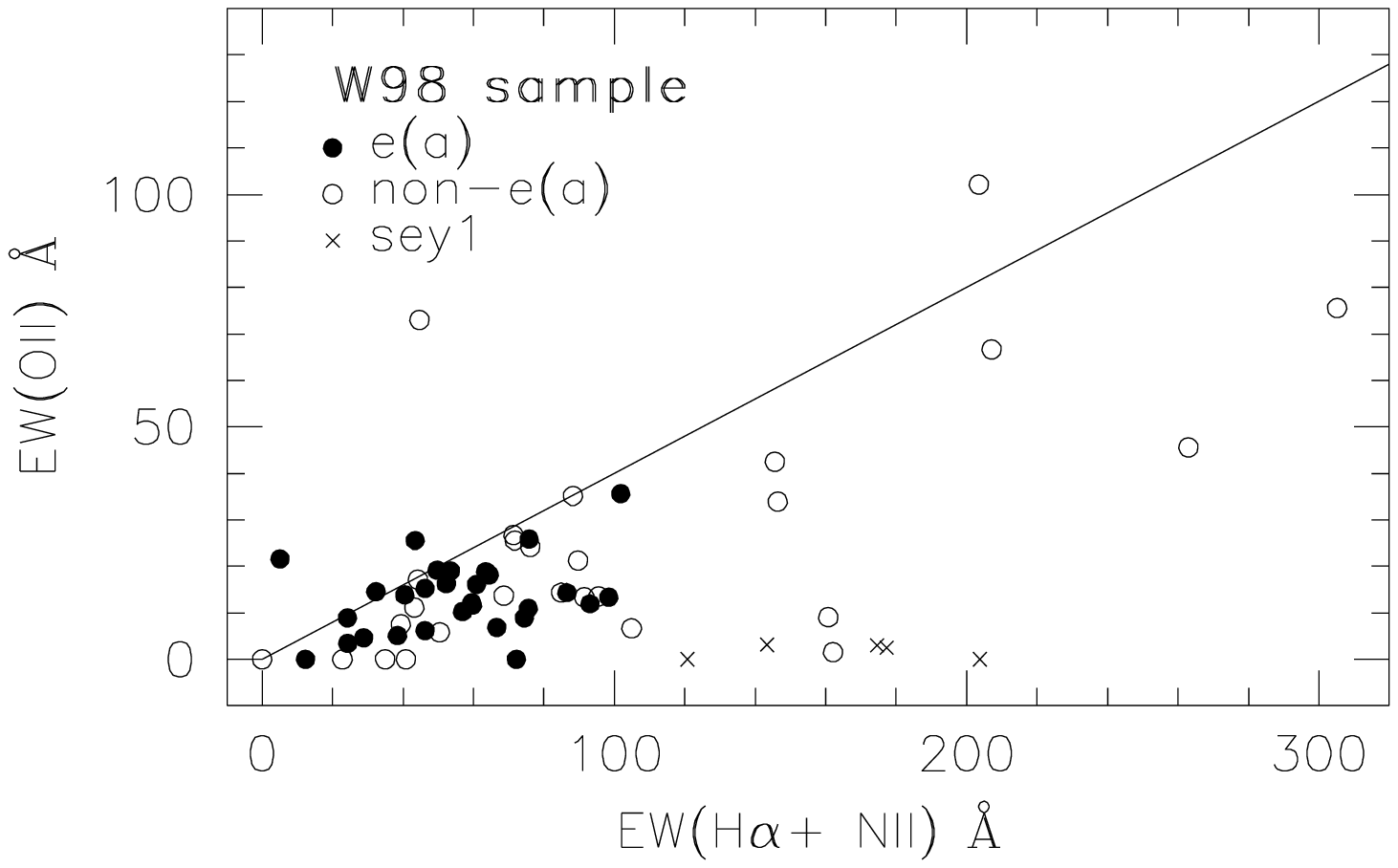,angle=0,width=4.0in}}
\vspace{-0.5in}

\noindent{\scriptsize
\addtolength{\baselineskip}{-3pt} 
\hspace*{0.1cm} Fig.~3.\ The line represents the 
fit for normal field galaxies in the local Universe
(EW([O{\sc ii}])=0.4 EW($\rm H\alpha$ + NII)). See P99, Fig.~5, for a
similar plot of the LK95 sample.
\addtolength{\baselineskip}{3pt}
}

\begin{table*}
{\scriptsize
\begin{center}
\centerline{\sc Table 5}
\vspace{0.1cm}
\centerline{\sc Median E(B-V) as a 
function of the spectral class}
\vspace{0.3cm}
\begin{tabular}{lccr}
\hline
\noalign{\smallskip}
Class & Median  & SD & N \\
      & E(B-V)  &    &   \\
\hline
\noalign{\smallskip}
e(a) & 1.11 & 0.12 & 34 \\
e(c) & 0.68 & 0.13 & 18 \\
e(b) & 0.62 & 0.15 &  5 \\
e    & 0.80 & 0.11 & 36 \\
\noalign{\smallskip}
all  & 0.86 & 0.07 & 93 \\
\noalign{\smallskip}
\noalign{\hrule}
\noalign{\smallskip}
\end{tabular}
\end{center}
}
\vspace*{-0.8cm}
\end{table*}

We now want to establish whether this hypothesis 
is in accordance with the values of the color excess E(B-V) derived by W98b
from the Balmer decrement $\rm H\alpha$/$\rm H\beta$. 
Table 5 lists the median E(B-V) and the bootstrap standard deviations 
of the main emission--line classes in the W98 sample. 
\footnote{Seyfert1 spectra have a very low average E(B-V) (W98b)
and have been excluded from this analysis.} 
We note that the e(a) galaxies appear to have a median E(B-V) significantly 
higher than the other classes and seem to 
represent the most extincted cases in a sample
which on average is already very dusty. 
\footnote{The E(B-V) differences among the spectral classes could 
be affected by the uncertainty in the measurement of the
underlying line in absorption, however we do not expect any systematic error 
that could cause an overestimate of the E(B-V) values for e(a)'s:
in the spectra with obvious $\rm H\beta$ absorption, the $\rm H\beta$
flux in emission has been corrected by W98a
performing a multi-component fitting, adopting a Lorentz profile in 
absorption and one or two Gaussian emission components. 
If anything, the absorption component is more likely to be unmeasured
in the weakest cases (non-e(a) spectra), with a consequent overestimate
of E(B-V) resulting in a trend contrary to what is observed.}  

In order to verify whether these E(B-V) values can account for the 
low [O{\sc ii}]/$\rm H\alpha$, we first need to determine the {\it intrinsic}
(unreddened) typical ratio. We will assume this to be equal to the \it 
intrinsic \rm ratio in normal nearby galaxies, which can be derived from
the median \it observed \rm ratio in spirals (0.43) correcting for the average 
extinction (1 mag at $\rm H\alpha$, K92). The correction in the case of
normal spirals therefore corresponds to a factor 2.5 for the
$\rm H\alpha$ flux and a factor about 5 for the [O{\sc ii}] flux if we 
assume a foreground screen dust configuration and we adopt 
the standard Galactic extinction curve for the diffuse medium
($R_V=A_V/E(B-V) =3.1$, Mathis 1990).
Combining the intrinsic typical ratio 
[O{\sc ii}]/${\rm H\alpha}_{int}$ ($\sim 0.85$) with the median E(B-V)
of our sample (0.86, Table 5), we expect to observe

[O{\sc ii}]/${\rm H\alpha}_{expec.}$=[O{\sc ii}]/${\rm H\alpha}_{int}$$*0.22$
= 0.19. 

This is broadly consistent with the median value observed in W98: 

[O{\sc ii}]/${\rm H\alpha}_{obs.}$=[O{\sc ii}]/${\rm H\alpha}_{int}$$*0.27$ =
0.23. 

We stress that there are two arbitrary assumptions in the method used above,
namely that the intrinsic line ratio in VLIRGs is equal to the 
one in normal spirals -- when the latter is 
corrected for the average extinction --
and that we can adopt as attenuation curve the 
extinction curve of the diffuse medium in the Galaxy. 
\footnote{Adopting as {\it attenuation curve}\
any of the {\it extinction curves} observed in our or nearby galaxies
is known to be incorrect in most cases because while the extinction curve
depends only on the physical properties of the dust grains,
the attenuation curve is determined also by the spatial distribution of 
stars and gas.}
Neverthless, the high extinction derived from the Balmer decrement
(E(B-V)) necessarily implies that the emission line ratios must be greatly 
affected by reddening and the comparison shown above demonstrates that 
the observed E(B-V) values are {\it consistent} with the low 
[O{\sc ii}]/$\rm H\alpha$.

\subsection{AGN types}

On the basis of the standard diagnostic diagrams discussed by
Veilleux \& Osterbrock (1987), W98b classified the spectra
into one of the following types (hereafter ``AGN type''): 
HII galaxy (H), LINER (L), mixture type between HII galaxy and LINER 
(LH), Seyfert2 (S2),
Seyfert1 (S1), unclassified AGN (S?), unclassified because lacking 
emission lines (O). 

The AGN type distribution of each spectral class is shown in Fig.~4.
The great majority of e(a)'s
are classified as HII galaxies or mixture types between HII galaxy and LINER
and only 1 e(a) galaxy has a Seyfert2 spectrum, which is a dubious case
between a Seyfert2 and a LINER. 
\footnote{Spectropolarimetric studies indicate that 
e(a) galaxies do not usually host obscured AGNs, since 
none of the approximately two dozen hidden broad--line regions detected to date
are found in objects dominated by a strong Balmer absorption--line
spectrum (Tran et al. 1998).} 
A similar AGN distribution is found for the e(c) spectra,
while only the e(b) class includes a higher proportion of Seyfert2 galaxies.
We conclude that e(a) VLIRGs do not generally display
spectral evidence for an AGN and that the only type of 
``activity'' observed in most of these e(a)'s is due to star formation.

\hbox{~}
\vspace{-0.5in}
\centerline{\psfig{file=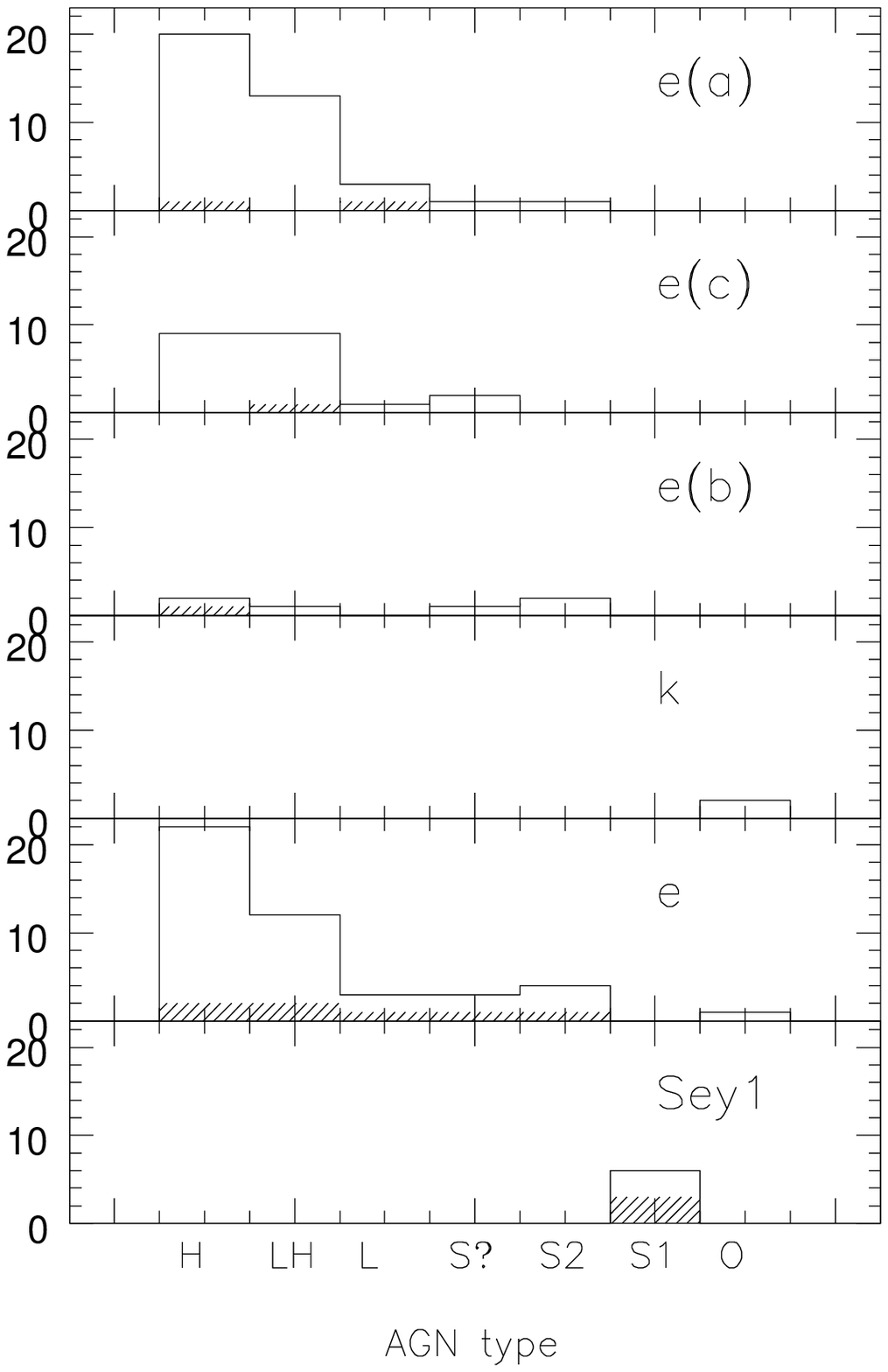,angle=0,width=4.0in}}

\noindent{\scriptsize
\addtolength{\baselineskip}{-3pt} 
\hspace*{0.1cm} Fig.~4.\ Histograms showing the distribution of AGN 
types within each spectral class.
The shaded area indicates galaxies with log $L_{FIR}/L_{\odot} > 12$.

\addtolength{\baselineskip}{3pt}
}

\section{The nature of the e(a) galaxies}

In this section we compare our spectroscopic results with 
the FIR and morphological properties with the aim of studying 
the star formation rates and the merger stage of the
e(a) galaxies. An interesting aspect of the comparison of the FIR and 
$\rm H \alpha$ luminosities is the determination
of the amount of star formation that cannot be detected at optical
wavelengths, while the analysis of the morphological properties of the e(a)'s 
can reveal whether they are associated with a specific phase of the merger
or with a typical separation between the two merging galaxies.

\subsection{Star formation rates}

The FIR and $\rm H\alpha$ luminosities are useful indicators of
the current star formation rate in dust--enshrowded and dust--free
galaxies respectively. 

The FIR luminosity distribution of the e(a) galaxies is similar to the one of
the whole VLIRG sample (Fig.~5), as shown by a ${\chi}^2$ test
at the 95\% confidence level. The median log$L_{FIR}(L_{\odot})$
is 11.68 for e(a)'s and 11.72 for the whole sample. 
The slightly higher value found for the
e(b)'s (11.87) is significant and is probably due to the fact that
the proportion of Seyfert galaxies in this spectral class is higher
than in the other classes. Seyfert1's are known to have the highest 
median FIR luminosity (12.64).

\hbox{~}
\vspace{-0.5in}
\centerline{\psfig{file=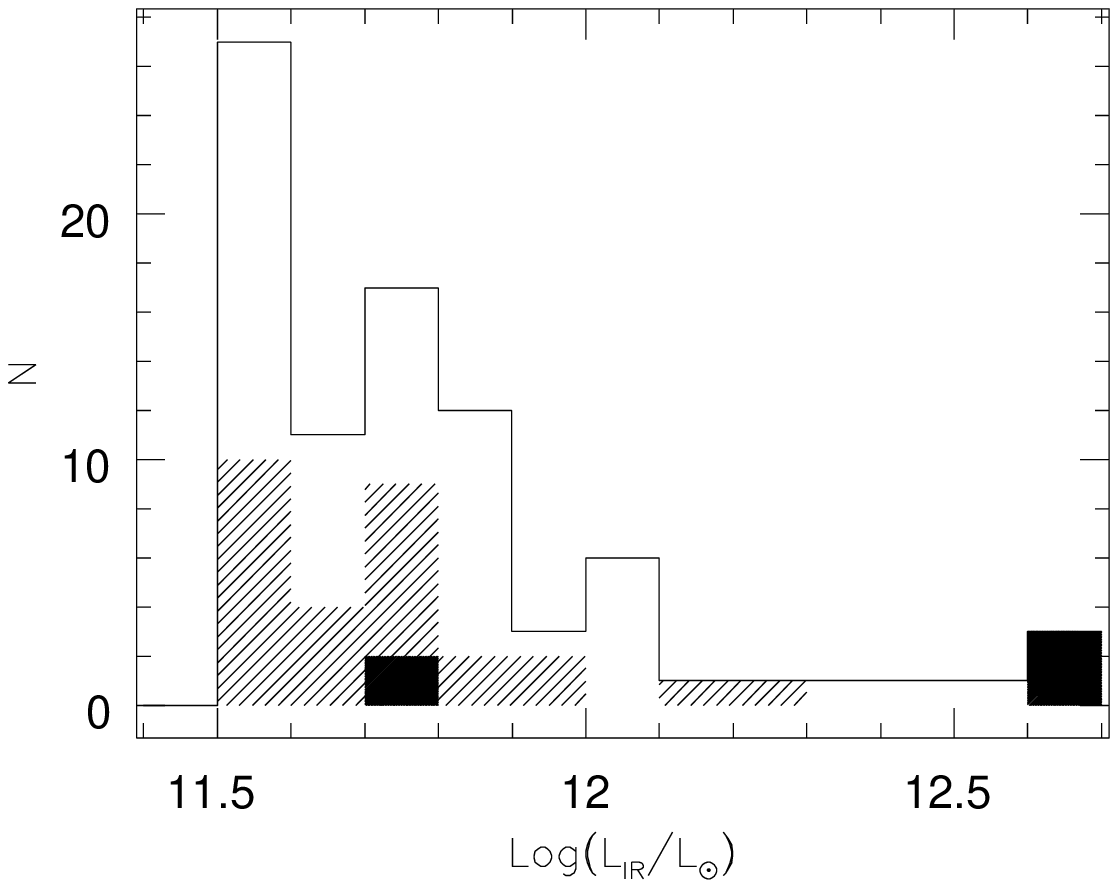,angle=0,width=4.0in}}
\vspace{-0.5in}

\noindent{\scriptsize
\addtolength{\baselineskip}{-3pt} 
\hspace*{0.1cm} Fig.~5\ FIR luminosity distribution of the
whole VLIRG sample (empty histogram), of the e(a) subsample (shaded histogram)
and of the Seyfert1 subsample (solid histogram). 

\addtolength{\baselineskip}{3pt}
}

In Fig.~6 we plot the FIR and $\rm H\alpha$ luminosities of the W98 sample.
The $\rm H\alpha$ luminosities have been derived from the
fluxes and the distances given in W98a,b.
The solid line is the best fit to the datapoints; 
for comparison we have plotted as a dotted line the relation valid for 
normal spiral galaxies at low--redshift (Devereux \& Young 1990),
which typically have $log \, L_{FIR}(L_{\odot})<10.5$.

The reader should keep in mind that our $\rm H\alpha$ luminosities
are derived from \it nuclear \rm spectra while the relation
for normal spirals was obtained from large aperture photometry.
The fractional contribution of nuclear emission to the total 
Halpha luminosities of interacting galaxies is three times higher 
than the average in isolated spirals 
(Kennicutt et al. 1987, K87): the median fraction is 13 \% and in most cases
is smaller/equal to 25 \%. Using K87 results and considering that:
a) the apertures in W98 are comparable to K87, but the median redshift
is higher (0.0324 versus 0.006/0.010);
b) in Infrared Luminous mergers the star formation is centrally concentrated
and it is reasonable to expect an even higher fractional
contribution from the nucleus (Kennicutt 1999, private communication);
then a correction factor of 7 in flux can be regarded as an
upper limit for the W98 galaxies and it is likely to overestimate 
the total luminosity.
The datapoints for integrated luminosities therefore should lie
in Fig.~6
between the solid line (best fit to the data, uncorrected) and the dashed line 
(L($\rm H\alpha$)$*7$). 

Even taking the aperture effects into account, the $\rm H\alpha $
luminosities at a given $L_{FIR}$ are much lower in the W98 galaxies
than in optically selected spirals, a result which is found in
Infrared Luminous samples in general. Considering only a subset of 
the data taken during the most favourable weather conditions does 
not alter this result.

\hbox{~}
\vspace{-0.5in}
\centerline{\psfig{file=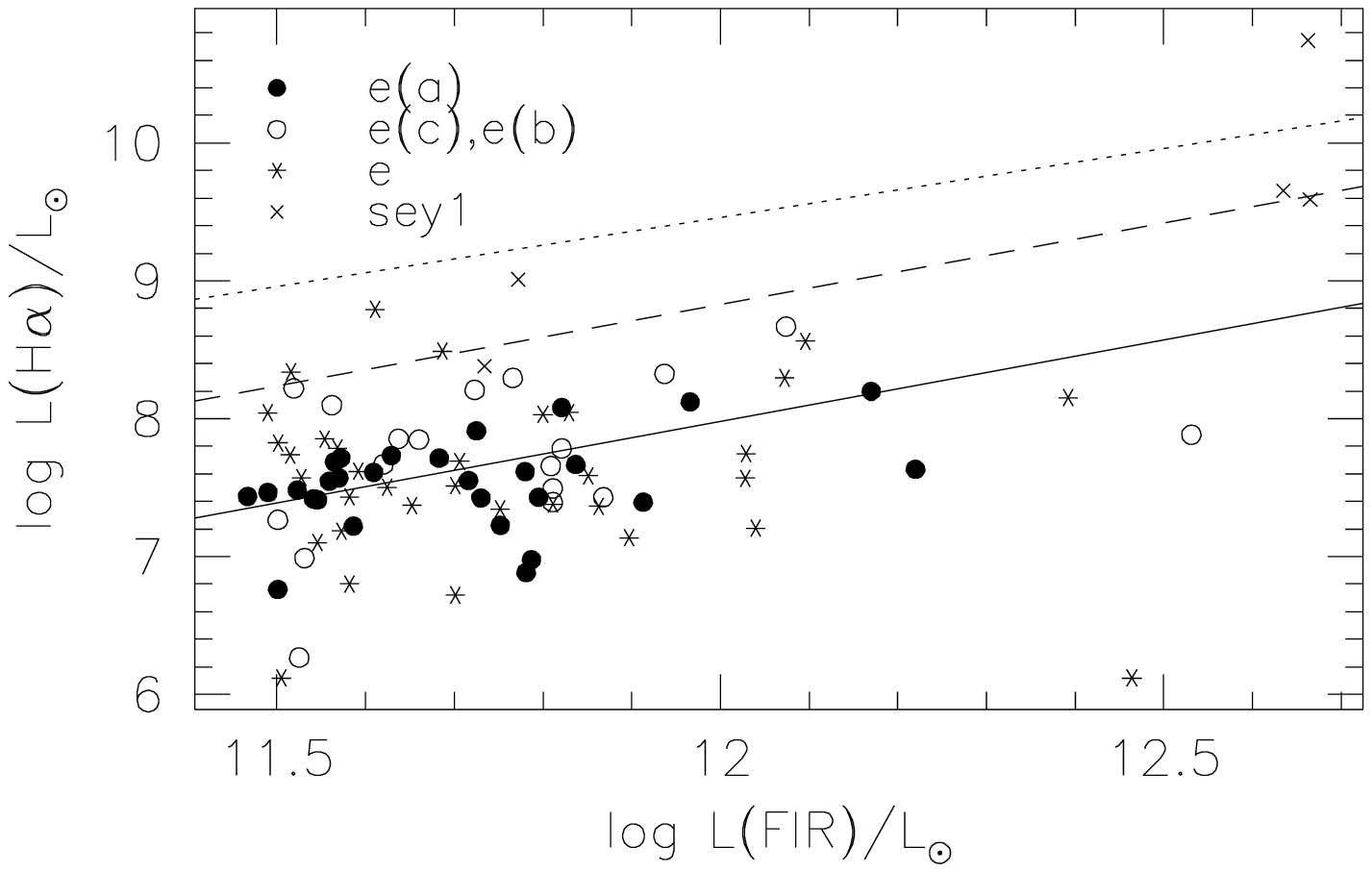,angle=0,width=4.0in}}
\vspace{-0.5in}

\noindent{\scriptsize
\addtolength{\baselineskip}{-3pt} 
\hspace*{0.1cm} Fig.~6\ FIR and $\rm H\alpha$ luminosities in solar units.
The best fit to the datapoints is shown as a solid line:
$\rm log \, L(H\alpha)(L_{\odot}) = 1.18 \times log L_{FIR}(L_{\odot}) - 6.18$.
The fit to the e(a) subsample is flatter:
$\rm log \, L(H\alpha)(L_{\odot}) = 0.67 \times log L_{FIR}(L_{\odot}) - 0.28$.
The relation found by Devereux \& Young (1990, DY90) for a sample of 
field spiral galaxies in the local Universe 
is extrapolated to the FIR luminosities of the present dataset 
and is shown as a dotted line
($L(\rm H\alpha)(L_{\odot}) \sim 3 \times 10^{-3} L_{FIR}(L_{\odot})$), 
after taking into account the correction for the different spectral FIR range.
The dashed line is found from the solid line multiplying $L_{\rm H\alpha}$
by 7 (see text) and represents the upper limit of the {\it integrated}
luminosity.

\addtolength{\baselineskip}{3pt}
}

This ``scarcity'' of $\rm H\alpha$ flux is directly mirrored in the
SFR estimates, as we will now show.
There are many calibrations of the $\rm H\alpha$ luminosity versus
the SFR (see Kennicutt 1998 for a review) and they are all found
using an evolutionary synthesis model assuming a Case B recombination
for ionization bound regions. Here we use the calibration given by 
Barbaro \& Poggianti (1997, BP97)
\footnote{The original calibration from BP97 included a factor 0.7
accounting for a fraction of the ionizing photons which is lost due to 
dust \it before \rm ionizing the gas
(see Mayya \& Prabhu 1996 for an observational confirmation of this effect). 
This factor is usually neglected in the other calibrations found in the 
literature and it has not been included here.}
for a Salpeter IMF in the mass range 0.1--100 $M_{\odot}$:
$SFR_{\rm H\alpha} \, (M_{\odot} \, yr^{-1})=1.1 
\times 10^{-41} L(\rm H\alpha$) (ergs $s^{-1}$).
\footnote{With the same IMF and based on an 
independent synthesis model, Kennicutt (1992b)
obtains: SFR ($M_{\odot} \, yr^{-1})=0.89 \times 10^{-41} L(\rm H\alpha$) 
(ergs $s^{-1}$), and Kennicutt (1998) gives:
SFR ($M_{\odot} \, yr^{-1})=0.79 \times 10^{-41} L(\rm H\alpha$) (ergs 
$s^{-1}$).}

For estimating the SFR from the FIR luminosity
we use the relation given by Kennicutt (1998):
$SFR_{FIR} \, (M_{\odot} \, yr^{-1})= 4.5 \times 10^{-44} L_{FIR} (ergs \,
s^{-1})$.
This is applicable only if most of the bolometric luminosity is
reprocessed in the FIR, as in the case of the sample considered here,
and for continuous bursts with ages less than $10^8$ years.
This latter assumption on the burst duration has not been ascertained for e(a)
galaxies, but the adopted SFR calibration 
lies within $\pm 30$ \% of most of the other published calibrations
and has comparable uncertainty.
Since both the $\rm H\alpha$ and the FIR calibrations
are obtained for the same IMF ad 
are sensitive to the hottest (more massive) young stars,
the ratio of the SFRs derived with the two methods is essentially 
independent from the IMF adopted. Furthermore, 
the ratio of SFRs is also distance-independent.

The SFRs derived with the two methods are presented in Table 2 and in Fig.~7.
The observed relation for e(a)'s is even flatter then for e(c)'s/e(b)'s
and corresponds to $SFR_{\rm H\alpha} \sim SFR_{FIR}*C_{ap}/70$,
where $1<C_{ap}<7$ is the aperture correction to be applied
to the $\rm H\alpha$ flux.
The errorbars in this relation are dominated by the uncertainties
in the conversion between fluxes and SFR, at least 30 \% 
both for the FIR and the $\rm H\alpha$, and in some cases the FIR 
luminosity is likely to include the emission from both companions.

The median E(B-V) of e(a) galaxies (Table 5)
corresponds to an extinction of 2.9 mag at the wavelength of
$\rm H\alpha$, if the standard
galactic extinction curve for the diffuse interstellar medium
is adopted (see \S3 for a discussion of the uncertainties connected to
this assumption). 
Neglecting aperture corrections to $\rm H\alpha$, this amount of reddening 
is not enough to reconcile the $SFR_{\rm H\alpha}$ estimate with the 
SFR derived from the FIR (dashed line in Fig.~7), while considering both
reddening and aperture effects, the two SFR estimates for e(a)'s 
are in agreement if the upper limit on the aperture correction is used (7). 
Adopting a more plausible smaller correction, the discrepancy between
the two SFRs could indicate
that a significant fraction of the FIR emission comes from a galactic region 
whose line flux is completely extincted.

From the comparison of FIR and $\rm H\alpha$ estimates of the SFR,
we conclude that the SFR derived from
the \it observed integrated \rm $\rm H\alpha$ flux 
in e(a) galaxies is a factor between 10 and 70
lower than the SFR derived from the FIR emission, depending on the
appropriate aperture correction to be applied to our sample.
Given the low [O{\sc ii}]/$\rm H\alpha$ ratios of e(a)'s, using the 
[O{\sc ii}] flux would underestimate the $SFR_{FIR}$ by a further factor of 2,
i.e. 20--140 in total.

\hbox{~}
\vspace{-0.5in}
\centerline{\psfig{file=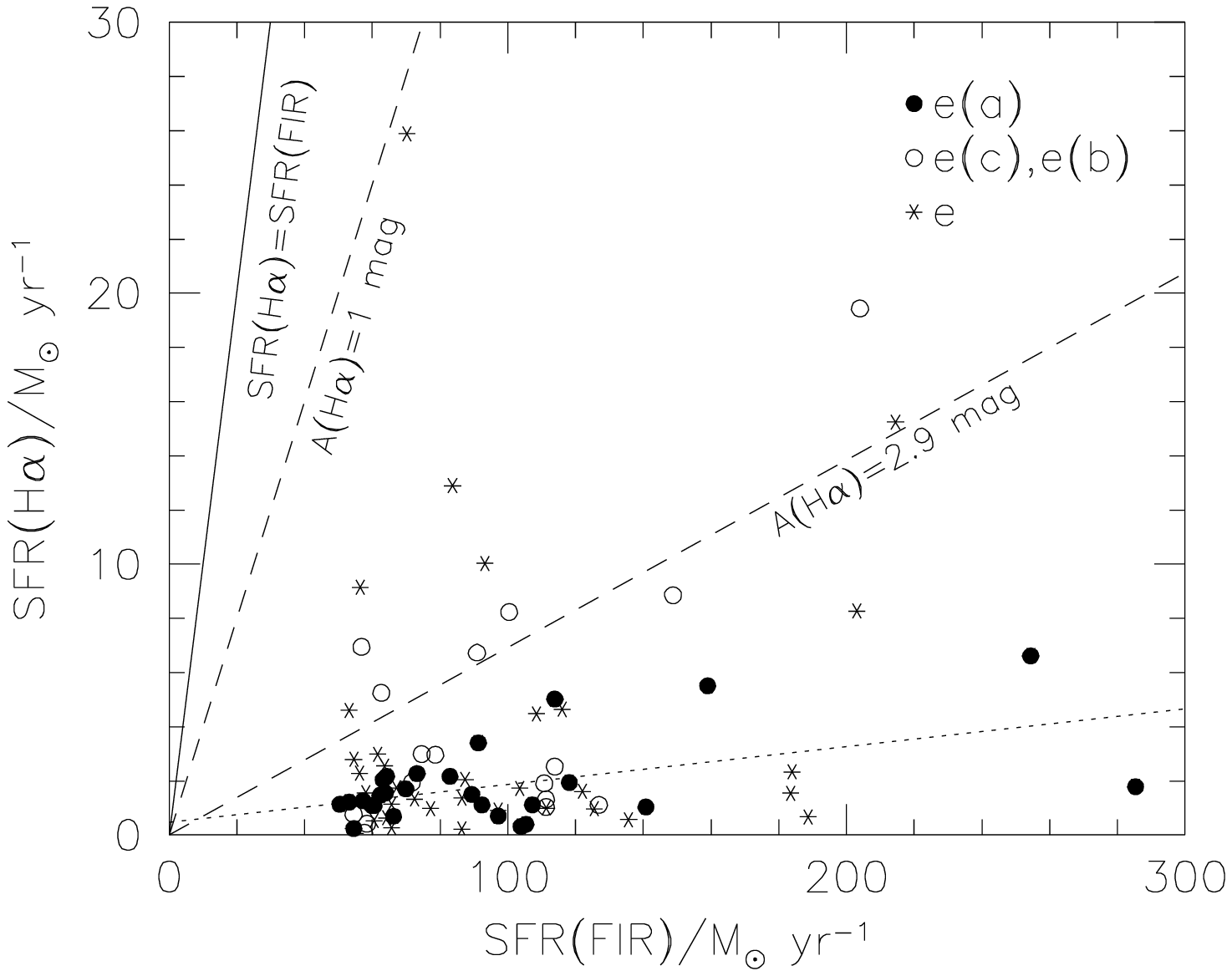,angle=0,width=4.0in}}
\vspace{-0.5in}

\noindent{\scriptsize
\addtolength{\baselineskip}{-3pt} 
\hspace*{0.1cm} Fig.~7\ 
The SFR derived from the $\rm H\alpha$ line versus the FIR--based estimate
(see text for details).
\it In this plot no aperture correction is applied to the $\rm H\alpha$
flux. \rm The dotted line is the fit to \it the e(a) population. \rm
The solid line shows the relation $SFR_{\rm H\alpha}=SFR_{FIR}$,
and the dashed lines are found for 1 mag extinction
at $\rm H\alpha$ (average extinction in K92 nearby spirals)
and for the 
average extinction in e(a)'s in W98 sample as determined by E(B-V) 
(A($\rm H\alpha)=2.9$ mag).
The X and Y ranges have been limited 
for displaying convenience, leaving 7 points out of the diagram. 
\addtolength{\baselineskip}{3pt}
}

\subsection{Morphological properties}

FIR luminous galaxies often show strong interaction or merging signatures
(Sanders \& Mirabel 1996, Duc et al. 1997);
based on optical images from the Digital Sky Survey, W98b assigned a 
morphological class
along a merger sequence (far to near companions 1--4, interacting pairs and
mergers 5--6, isolated/advanced mergers 0). 
The ``0'' class should include both
truly isolated galaxies and objects in an advanced or post merging stage.
	
In the VLIRG sample the proportion of close mergers is very high for
all the spectral classes (Table 6). 
The ambiguity in interpreting the 0 class complicates the comparison:
if the tendency of the e(c)'s to 
avoid the 5-6 class is real, then maybe a large fraction of the e(c) 
galaxies of type ``0'' are truly isolated objects, while the ``0'' type
Seyfert1 are likely to be in an advanced merger stage. 
The e(a)'s are mostly merging galaxies/interacting pairs
or galaxies with close companions. The median \it projected \rm 
separation between  the e(a) galaxies and their companions
is 50 kpc and the highest separations observed are 100--150 kpc (W98b).

\begin{table*}
{\scriptsize
\begin{center}
\centerline{\sc Table 6}
\vspace{0.1cm}
\centerline{\sc Fraction of VLIRGs of a given spectral class as a function
of merging type}
\vspace{0.3cm}
\begin{tabular}{lccccr}
\hline
\noalign{\smallskip}
Class & 0 & 1--2 & 3--4 & 5--6 & N \\
\hline
\noalign{\smallskip}
e(a) & 0.14$\pm$0.07  & 0.10$\pm$0.06 & 0.21$\pm$0.08 &  0.55$\pm$0.14 & 29 \\
e(c) & 0.46$\pm$0.19  & 0             & 0.15$\pm$0.11 &  0.38$\pm$0.17 & 13 \\
e(b) & 0              & 0             & 0.20$\pm$0.20 &  0.80$\pm$0.40 &  5 \\
sey1 & 0.20$\pm$0.20  & 0             & 0.20$\pm$0.20 &  0.60$\pm$0.35 &  5 \\
e    & 0.07$\pm$0.05  & 0.13$\pm$0.07 & 0.20$\pm$0.08 &  0.60$\pm$0.14 & 30 \\
\noalign{\smallskip} 
\noalign{\hrule}
\noalign{\smallskip}
\end{tabular}
\end{center}
}
\vspace*{-0.8cm}
\end{table*}

\section{E(a) galaxies in other spectroscopic surveys}

In this section we review the detection of e(a) galaxies in other
spectroscopy surveys at low and high redshift,
in order to identify the common properties shared by all e(a)'s.
What follows is not a complete inventory (see P99 for 
additional references and van Breugel 1999 for radio selected e(a)'s) and
we will focus on those observations which can be useful for
understanding how the e(a) population evolves with redshift and 
depends on the environment.

\subsection{E(a) galaxies in the local Universe}

At low-$z$, e(a) spectra are present but scarce in optically--selected 
spectroscopic surveys of field galaxies:
P99 noticed that at most 7\% (1/14) of the normal
non--merging, non--Seyfert galaxies in K92\footnote{The e(a) fraction
in K92a cannot be considered representative of a complete sample, 
since this survey is not complete in any sense and consists of nearby galaxies
covering a broad range in morphological types, luminosities and star formation
properties.} show an e(a) spectrum.
Hashimoto (1998, private communication) 
estimated the e(a) frequency to be 8 \% in the Las Campanas Redshift Survey
(LCRS, see also Fig.~1 in Zabludoff et al. 1996), but the comparison with 
the W98 sample is uncertain owing to the differences in
the classification criteria and in the spectral quality.
An analysis of the FIR and morphological properties of the e(a)'s
in the LCRS has not yet been undertaken.

As discussed in P99, a high proportion of e(a) spectra is found instead
in the LK95 sample: these galaxies were selected for being merging or 
strongly interacting systems, either in the optical or in the near--infrared,
and are all strong FIR emitters. 
Evidence for a high e(a) incidence is also found 
in an unusually active compact group of galaxies
with strong FIR and CO emission (de Carvalho \& Coziol 
1999). 
\footnote{At least 14 \% of our VLIRGs are in groups or
multiple--merging systems. An evolutionary connection between 
ULIRGs/VLIRGs and compact groups has been suggested by W98b.}

The analysis of the W98 sample of Very Luminous Infrared Galaxies
and of the other ``e(a)-rich'' samples seems to indicate that field/group e(a) 
galaxies in the local Universe are 
 highly extincted starbursts 
and that they are preferentially merging/strongly interacting systems. 
A statistical study of the e(a) occurrence in nearby clusters is not yet 
available, but we know it must be rare, given the small number of 
emission-line galaxies present in dense environments at low-$z$.

\subsection{E(a) galaxies at high redshift}

The relevance of e(a) spectra for galaxy evolution studies was first
discussed by D99/P99 who found that e(a)'s constitute 
about 10 \% of their sample of cluster and field galaxies
at $z \sim 0.4-0.5$.
HST images of e(a) galaxies in these clusters show that at least half of
them are involved in a merger or strong interaction (D99, P99).
Spectra with e(a) characteristics 
were previously observed in other spectroscopic surveys of 
distant clusters
(Couch \& Sharples 1987, Couch et al. 1994, 1998, Fisher et al. 1998).

At redshifts comparable to the D99 catalog and up to $z=1$, a population 
of Balmer strong galaxies with emission is present in  the field sample of the
Canada-France Redshift Survey (Hammer et al.\ 1997, H97,
see P99 for a discussion).
More recently Flores et al. (1998a,b) presented the results of a deep
ISO survey of one of the CFRS fields at 6.75 and 15$\micron$.
At 6.75$\micron$ 8 out of the 15 detected galaxies have e(a) spectra
(EW($\rm H\delta) \gs 5$ \AA $\,$ in this case) and lie at $0.07<z<0.88$;
the remaining ISO sources have Seyfert1/Seyfert2/e(b) spectra. 
Among other 8 less secure 6.75$\micron$ sources, 6 display an e(a) spectrum.
At 15$\micron$ most (71 \%) of the sources with optical spectroscopy 
are classified as e(a) galaxies and have a median redshift $<z>=0.76$.
The FIR luminosities of the Flores et al. e(a)'s span the range 10.76 to
12.31 log$L/L_{\odot}$, with a median 11.66
and their moderate EW(O{\sc ii}) are comparable to those in W98 spectra.
This is the first confirmation of the IR luminous nature of e(a) galaxies
at high redshift, except for the sub--mm detection 
of a distant e(a) cluster galaxy, \#834 in Cl\,0024$+$16
(Smail et al.\ 1998a). Preliminary results of other field ISO surveys
seem to reinforce this conclusion (Elbaz et al. 1999); in distant clusters,
Duc et al. (1999) detect at 15$\micron$ all the e(a) galaxies in the cluster 
A1689 ($z \sim 0.2$) and confirm their dusty nature and high star
formation rates.

The e(a) phenomenon is obviously widespread both al low and high redshifts
and involves a high fraction of galaxies
in all the FIR luminous samples at any $z$ studied so far.
In the distant Universe e(a)'s are numerous both in the field
and in the cluster environment; a systematic census at low-$z$
will be needed to assess the amount of evolution of the e(a) population
and its dependence on the environment.

\section{Discussion}

We have seen in \S3 that e(a) spectra are exceptionally numerous among VLIRGs,
but their incidence in the W98 sample (Table 4) is based
on nuclear spectra, while distant galaxy spectra should be more
appropriately compared with 
\it integrated spectra \rm of low-redshift galaxies. 
We can evaluate the relevance of aperture effects from the 
catalog of LK95 which gives both nuclear and whole aperture/integrated spectra
of the merging galaxies: in 3 out of the 16 e(a)'s the galaxy has an e(a) 
nuclear spectrum and a different whole aperture/integrated
spectral type, suggesting that a spectroscopic survey of distant galaxies
could miss about 20 \% of the ``nuclear'' e(a) population.

In \S3 we also noted that no k+a's are found in our sample.
This suggests that the k+a/a+k galaxies are not very luminous FIR emitters
and this result
is consistent with the lack of emission lines in their optical spectrum
and with the \it post \rm --starburst interpretation.
Zabludoff et al. (1996) found that at least 5 of the 21
low--redshift ``E+A's'' (= k+a/a+k according to our definition) galaxies 
in the Las Campanas Redshift Survey have clear tidal features.
The absence of this type of spectra in the W98 sample 
implies that if a merger/strong interaction is the mechanism responsible
for the field k+a/a+k galaxies at low redshift,
they are probably observed during an evolved (post) merger stage, as 
Zabludoff et al. suggest.

We would like to stress here that \it by definition \rm 
k+a (``E+A'') spectra
have \it no detectable emission lines \rm (Dressler \& Gunn 1983),
at least at the level of the [O{\sc ii}] detection limit of the 
available spectroscopic surveys of distant galaxies (typically 3 \AA). 
Unfortunately in the literature 
this notation has sometimes been used for emission--line galaxies. 
This work demonstrates once more that
the presence/lack of emission lines is not merely a ``spectral detail'', but
reveals a current/absent
star formation activity and discriminates between
post--starburst and dusty starburst galaxies. The possibility of an 
evolutionary connection between k+a's and e(a)'s has been discussed in P99,
who also noted that some of the k+a spectra might
be extreme cases of e(a) galaxies, in which the [O{\sc ii}] line is totally
obscured by dust.

\subsection{The physical origin of the e(a) spectra}

In the following we want to investigate the physical origin of the
e(a) spectral characteristics, namely
a) the strong $\rm H\delta$ line in absorption and b) the low 
[O{\sc ii}]/$\rm H\alpha$ ratio.

On the basis of dust--free models (P99), e(a) spectra can be interpreted
as {\it post--starburst} galaxies with some residual star formation
\footnote{A similar interpretation is proposed by Flores et al. 1998b, 
who found
their ISO data to be consistent with the scenario of strong starburst episodes 
followed by the last phases of the burst, where the IR emission is still 
high due to dust heating by intermediate--mass stars (1--3 $M_{\odot}$).},
but the analysis of W98 and other samples (\S5) shows that 
whenever the discriminating data (IR) are available,
e(a) galaxies are found 
to be starbursts which contain a large amount of dust, 
\rm hence we cannot safely rely on dust--free 
models to estimate their SFRs and histories. 

As discussed in \S4.1, the low [O{\sc ii}]/$\rm H\alpha$ ratios are probably
due to the high dust extinction which affects the emission line fluxes --
generated inside the HII regions -- much more than the stellar 
underlying continuum coming from populations that are more widely
distributed throughout the galaxy (Calzetti et al. 1994, Mayya \& Prabhu 1996).
In principle other factors influence [O{\sc ii}]/$\rm H\alpha$, such as
metallicity, \footnote{The line ratio 
is expected to decrease with increasing metallicity for 
an HII region with $Z$ higher than solar (Stasinska 1990), but
an analysis of the $R_{23}$ indices (Zaritsky et al. 1994) of W98 galaxies 
rules out exceptionally high abundances.}
or high ionization,\footnote{High ionization is ruled out 
from the observed [O{\sc iii}]/$\rm H\beta$ (W98b).},
but we have shown that the most probable cause of the
observed ratios is the high extinction which is implied by
the E(B-V) values derived from the Balmer decrement.

A selective extinction is also capable of explaining the strong $\rm H\delta$
line in absorption, which mostly arises from stars with
lower mass/effective temperature (7000-14000 K) than the O, B stars 
that ionize the gas ($> 25000$ K).
The stellar generations  
responsible for the strong $\rm H\delta$ line, with ages between a few 
$10^7$ and $1.5 \, 10^9$ yrs 
(Poggianti \& Barbaro 1996, 1997), have had the time
to drift away from (or disperse) the dusty molecular clouds where they were
born (Calzetti et al., 1994 and references therein) and their emission
dominates the integrated spectrum at $\lambda \sim 4000$ \AA $\,$ if younger
stars (causing the emission filling and with lower
stellar EW($\rm H\delta$)) are more heavily obscured.
Observing an EW($\rm H \delta)> 4$ \AA $\,$ therefore suggests
that the on-going starburst started at least $4-5 \, 10^7$ yrs before the time
of the observation. 

A simple way to depict this scenario is to imagine that \it the extinction
towards a star/stellar population decreases with the stellar age \rm
(Kinney, private communication):
the effect of dust is maximum for the youngest generation of
stars (that provide the ionizing flux
responsible for the emission lines, timescale $<$ few times
$10^7$ yr), and decreases for older stars.
In a sense, this scenario is equivalent to a ``clumpy
distribution of dust'' where the location and thickness of the patches
are not random, but depend on the age of the embedded stellar populations.
The integrated emerging spectrum at wavelength $\lambda$ is hence dominated by 
the less extincted (older) stars that 
are strong emitters at that $\lambda$, 
or that are indirectly responsible for the flux $F_{\lambda}$ (for example 
for the nebular emission). 
\footnote{The FIR flux can be 
influenced by stars in a quite broad age range, but in starbursts
it is mostly affected by the youngest populations and the fact that a fraction
of the stars in starbursts seem to be embedded in optically thick dust clouds
(Calzetti et al. 1995) suggests that a non-negligible fraction of the FIR 
emission arises from the innermost layer of the dusty ``onions''.}
This happens because only the least obscured regions give the
major contribution to the integrated spectrum (Witt et al. 1992).

An age-dependent obscuration is expected to mark in a peculiar way
the spectrum of a dust-enshrouded starburst. 
If dust is absent, each given portion of a galactic spectrum
is due to the stellar populations of a specific range 
of ages: for example, the emission lines are only observed if young stars
with age $< 4 \, 10 ^7$ yrs are present, while in a star-forming
galaxy the U band light is influenced by stars of ages $< 10^9$ yrs.
Considering now the effects of dust,
if each stellar age corresponds to a different amount of obscuration, {\it
within the same spectrum} we expect to measure different values of extinction,
depending on the spectral region/feature used to estimate it.
For example, UV-based reddening measurements should underestimate
the extinction of the emission lines.

Observational evidence supports this scenario;
the extinction values derived by techniques using different
spectral ranges/features generally disagree 
(Fanelli et al. 1988, Bohlin et al. 1990, Keel 1993,
Calzetti et al. 1994, Veilleux et al. (1995), Lancon et al. 1996,
Mayya \& Prabhu 1996, Calzetti 1997,1998, Mas-Hesse \& Kunth 1999): 
the reddening 
of the UV/optical stellar continuum in starburst galaxy spectra is lower than
the reddening of the ionized gas, and this latter is lower than
the one inferred from the comparison of the Balmer fluxes with the 
nebular free--free radio emission (Israel \& Kennicutt 1980, Viallefond 
et al. 1982, Caplan \& Deharveng 1986). 

Mid/Far--IR and optical studies with good spatial resolution 
(see e.g. Hwang et al. 1999) hopefully will 
provide a means to investigate the spatial segregation 
of the various spectral components
and the selective nature of dust extinction in the integrated spectra of
starburst galaxies.

\section{Summary}

$\bullet$ The analysis of a complete sample of Very Luminous InfraRed Galaxies
confirms that emission--line spectra with a strong $\rm H\delta$ line in 
absorption (e(a) spectra) are numerous among FIR luminous galaxies, 
as suggested by P99.
More than 50 \% of the VLIR galaxies show such a spectrum, in contrast
with optically selected galaxy samples where the e(a) fraction is small,
but an accurate assessment of this small e(a) fraction awaits
a similar spectral analysis 
on a well-defined control sample of non-IRAS galaxies.
The incidence of e(a) galaxies in the different samples 
seems to suggest
that the e(a) signature is capable of identifying from optical data alone a
population of heavily extincted starburst galaxies, while neither the moderate
strength of their emission lines nor their optical colors (P99, Trentham et 
al. 1999, Elbaz et al. 1999) 
can distinguish them from galaxies with more modest rates of star 
formation. 
We note that these results do not rule out a viable alternative 
producing e(a) spectra, although no such case is known yet: 
a post-starburst galaxy with some residual star formation (P99).
A systematic study of optically-selected e(a) galaxies is required
to establish the occurrence of this type of e(a)'s.

$\bullet$
The median FIR luminosity of e(a) galaxies in our sample is log 
$L_{FIR}/L_{\odot} \le 11.68$, but we stress that
this should be regarded as an upper limit to
the whole FIR luminosity distribution of e(a)'s which
needs to be determined from the analysis of spectroscopic catalogs 
including galaxies with log $L_{FIR}/L_{\odot} < 11.5$
(Kim et al. 1995). While UltraLuminous InfraRed Galaxies 
can be found efficiently up to very high redshifts by SCUBA at sub--mm 
wavelengths, the e(a) signature may help to identify less extreme, but more 
common dust-enshrouded galaxies and it appears to be essentially free from AGN 
contamination. 

$\bullet$
A universal property of the e(a) spectra appears to be a low 
[O{\sc ii}]/$\rm H\alpha$ equivalent width ratio; we find this to be due to a 
correspondingly low flux ratio, which is equal to 1/2 of the value observed
in nearby spirals. We interpret this result as an effect of dust extinction,
which is consistent with the average color excess E(B-V) derived from the 
Balmer decrement. We propose a scenario of \it selective \rm dust obscuration
that affects the youngest stellar populations more than the older stars
and that can explain also the strong $\rm H\delta$ absorption lines observed.

$\bullet$
The star formation rates derived from the FIR luminosities of the e(a)
galaxies lie in the range 50-300 $M_{\odot} \, yr^{-1}$; 
SFR estimates based on the $\rm H\alpha$ (and [O{\sc ii}]) luminosities
yields values a factor of 10-70 (20-140) lower.

$\bullet$
The incidence of mergers/strong interactions is very high in Very
Luminous Infrared Galaxies of any spectral class and the only
notable difference among the classes is a higher proportion of
e(c) galaxies with no obvious sign of interaction with a companion,
which could be either advanced mergers or truly isolated objects.
About 75\% of our e(a) galaxies are mergers/interacting pairs with close 
companions.

$\bullet$
A significant population of e(a) galaxies has been found by deep optical 
spectroscopic surveys both in the cluster and the field environment at 
$z \sim 0.4-0.5$ (D99, P99, H97). In the low-$z$ Universe, a systematic 
census of e(a)'s is needed in order to assess the amount of evolution as 
a function of redshift and environment.

\section*{Acknowledgements} 

We wish to thank C. Liu and R. Kennicutt for providing us their
spectrophotometric atlas of merging galaxies and we acknowledge the 
availability of the Kennicutt's (1992a) atlas of  galaxies 
from the NDSS-DCA  Astronomical Data Center.  
This work has greatly benefited from discussions with
Ian Smail, Pierre-Alain Duc, Steve Maddox, Robert Kennicutt, Paola
Mazzei, Guido Barbaro and Fabio Governato, and has been supported in 
part by the Formation and Evolution of 
Galaxies network set up by the European Commission under contract 
ERB FMRX-CT96-086 of its TMR program.  
HW acknowledges support from the Royal Society China Royal Fellowship scheme.

\smallskip


\begin{thebibliography}{}
\itemsep=0in

\bibitem{}
Altieri, B., Metcalfe, L., Kneib, J.\,P., McBreen, B., Aussel, H., Biviano, A.,
Delaney, M., Elbaz, D., Leech, K., Lemonon, L., Okumura, K., Pello, R., 
Schulz, B., 1999, A\&A in press (astro-ph 9810480)

\bibitem{}
Aussel, H., Cesarsky, C.\,J., Elbaz, D., Starck, J.\,L., 1999, A\&A, 342, 313

\bibitem{}
Barbaro G. \& Poggianti B.\,M., 1997, A\&A, 490, 504

\bibitem{}
Barger, A., Cowie, L., Sanders, D., Fulton, E., Taniguchi, Y., Sato, Y.,
Kaware, K., Okuda, H., 1998, Nature, 394, 248

\bibitem{}
Bohlin, R.\,C., Cornett, R.\,H., Hill, J.\,K., Stecher, T.\,P., 1990, ApJ,
363, 154

\bibitem{}
Calzetti, D., Kinney, A.\,L., Storchi-Bergmann, T., 1994, ApJ, 429, 582

\bibitem{}
Calzetti, D., Bohlin, R.\,C., Kinney, A.\,L., Storchi-Bergmann, T., Heckman, 
T.\,M., 1995, ApJ, 443, 136

\bibitem{}
Calzetti, D., 1997, AnJ, 113, 162

\bibitem{}
Calzetti, D., 1998, in {\it Dwarf Galaxies and Cosmology}, eds. Thuan et al.,
in press 

\bibitem{}
Caplan, J., Deharveng, L., 1986, A\&A, 155, 297


\bibitem{}
Cimatti, A., Andreani, P., Rottgering, H., Tilanus, R., 1998, Nature, 392, 895

\bibitem{}
Clements, D.\,L., Desert, F-X., Franceschini, A., Reach, W.\,T., Baker, A.\,C.,Davies, J.\,K., Cesarsky, C., 1999, A\&A in press (astro-ph 9901267) 

\bibitem{}
Cohen, J.\,G., Hogg, D.\,W., Blandford, R., Pahre, M.\,A., Shopbell, P.\,L.,
in {\it Infrared Surveys: A Prelude to SIRTF}, in press (astro-ph 9808343)

\bibitem{}
Couch W.\,J., Sharples R.\,M., 1987, MNRAS, 229, 423 (CS87)

\bibitem{}
Couch, W.\,J., Ellis, R.\,S., Sharples, R.\,M., Smail, I., 1994, ApJ, 430,
 121

\bibitem{}
Couch W.\,J., Barger A.\,J., Smail, I., Ellis R.\,S., Sharples R.\,M., 1998,
ApJ, 497, 188

\bibitem{}
de Carvalho, R.\,R., \& Coziol R., 1999, preprint (astro-ph 9901006)

\bibitem{}
Devereux, N.\,A., Young, J.\,S., 1990, ApJ, 350, L25

\bibitem{}
Dey, A., Graham, J.\,R., Ivison, R.\,J., Smail, I., Wright, G.\,S., 
Liu, M., 1999, ApJ, submitted (astro-ph 9902044)

\bibitem{}
Dickinson, M., 1998, in {\it Hubble Deep Field}, eds. Livio et al., p. 219

\bibitem{}
Dressler A., Gunn J.\,E., 1983, ApJ, 270, 7

\bibitem{}
Dressler, A., Smail, I., Poggianti, B.\,M., Butcher, H., Couch, W.\,J.,
Ellis, R.\,S., Oemler, A.\ Jr., 1999, ApJS, in press (D99) (astro-ph
9901263)

\bibitem{}
Duc, P.\,A., Mirabel, I.\,F., Maza, J., 1997, A\&AS, 124, 533

\bibitem{}
Duc, P.\,A., Poggianti, B.\,M. et al, 1999, in preparation

\bibitem{}
Elbaz, D., Aussel, H., Cesarsky, C.\,J., Desert, F.\,X., Fadda, D., 
Franceschini, A., Puget, J.\,L., Starck, J.\,L., 1998, in
{\it Next Generation Space Telescope}, Proc. of 34th Liege Intern. Astroph.
Colloquium, in press

\bibitem{}
Elbaz, D., Aussel, H., Cesarsky, C.\,J., Desert, F.\,X., Fadda, D.,
Franceschini, A., Harwit, M., Puget, J.\,L., Starck, J.\,L.,  1999,
in {\it The Universe as Seen by ISO}, 
eds. Cox \& Kessler, in press
(astro-ph 9902229)

\bibitem{}
Ellis, R.\,S., 1998a, ARAA, 35, 389

\bibitem{}
Ellis, R.\,S., 1998b, Nature, 395, A3 (reviews)

\bibitem{}
Fanelli, M.\,N., O'Connell, R.\,W., Thuan, T.\,X., 1988, ApJ, 334, 665

\bibitem{}
Fisher, D., Fabricant, D., Franx, M., van Dokkum, P., 1998, ApJ, 498, 195

\bibitem{}
Flores, H., Hammer, F., Desert, F.\,X., Cesarsky, C., Thuan, T.\,X., 
Crampton, D., Eales, S., Le Fevre, O., Lilly, S.\,J., Elbaz, D., Omont, A., 
1998a, A\&A in press (astro-ph 9811201)

\bibitem{}
Flores, H., Hammer, F., Thuan, T.\,X., Cesarsky, C., Desert, F.\,X.,
Omont, A., Lilly, S.\,J., Eales, S., Crampton, D., Le Fevre, O.,
1998b, ApJ in press (astro-ph 9811202)


\bibitem{}
Glazebrook, K., Blake, C., Economou, F., Lilly, S., Colless, M.,
1999, MNRAS in press (astro-ph 9808276)

\bibitem{}
Graham, J.\,R., Dey, A., 1996, ApJ, 471, 720

\bibitem{}
Hammer, F., Flores, H., Lilly, S.\,J., Crampton, D., Le Fevre, O., Rola, C.,
Mallen-Ornelas, G., Schade, D., Tresse, L., 1997, ApJ, 481, 49 (H97)

\bibitem{}
Helou, G., Khan, I.\,R., Malek, L., Boehmer, L., 1988, ApJS, 68, 151

\bibitem{}
Hughes, D., Serjeant, S., Dunlop, J., et al., 1998, Nature, 394, 241

\bibitem{}
Hwang, C-Y., Lo, K.\,Y., Gao, Y., Gruendl, R.\,A., Lu, N., 1999,
ApJL, 511, L17

\bibitem{}
Israel, F.\,P., Kennicutt, R.\,C., 1980, Astrophys. Letters, 21, 1


\bibitem{}
Keel, W.\,C., 1993, in {\it Massive Stars: Their Lives in the Interstellar
Medium}, ASP Conf. Ser. vol. 35, p.498

\bibitem{}
Kennicutt, R.\,C.\ Jr, 1992a, ApJS, 79, 255

\bibitem{}
Kennicutt, R.\,C.\ Jr, 1992b, ApJ, 388, 310

\bibitem{}
Kennicutt, R.\,C.\,Jr, Keel, W.\,C., van der Hulst, J.\,M., Hummel, E.,
Roettiger, K.\,A., 1987, AnJ, 93, 1011

\bibitem{}
Kennicutt, R.\,C.\ Jr, 1998, ARAA, 36, 189


\bibitem{}
Kim, D.\,C., Sanders, D.\,B., Veilleux, S., Mazzarella, J.\,M.,
Soifer, B.\,T., 1995, ApJS, 98, 129

\bibitem{}
Lancon, A., Rocca-Volmerange, B., Thuan, T.\,X., 1996, A\&AS, 115, 253

\bibitem{}
Lilly, S.\,J., Eales, S.\,A., Gear, W.\,K.\,P., Hammer, F., Le Fevre, O.,
Crampton, D., Bond, J.\,R., Dunne, L., 1999, ApJ in press (astro-ph 9901047)

\bibitem{}
Liu, C.T., Kennicutt, R.\,C. Jr, 1995a, ApJS, 100, 325

\bibitem{}
Liu, C.T., Kennicutt, R.\,C. Jr, 1995b, ApJ, 450, 547

\bibitem{}
Mas-Hesse, J.\,M., Kunth, D., 1999, A\&A in press (astro-ph 9812072)

\bibitem{}
Mathis, J.\,S., 1990, ARAA, 28, 37

\bibitem{}
Mayya, Y.\,D., Prabhu, T.\,P., 1996, AnJ, 111, 1252

\bibitem{}
Metcalfe, L., Altieri, B., McBreen, B., Kneib, J-P., Delaney, M., Biviano, A.,
Kessler, M.\,F., Leech, K., Okumura, K., Schulz, B., Elbaz, D., Aussel, H.,
1999, in  \it The Universe as seen by ISO, \rm 
eds. Cox \& Kessler, in press, (astro-ph 9901147)

\bibitem{}
Meurer, G., Heckman, T.\,M., Lehnert, M.\,D., Leitherer, C., Lowenthal, J.,
1997, AnJ, 114, 54

\bibitem{}
Oliver, S., Rowan-Robinson, M., Cesarsky, C., et al., 1999, in {\it Wide-field
Surveys in Cosmology}, Proc., XIV IAP Meeting, in press (astro-ph 9901274)

\bibitem{}
Pettini, M., Kellogg, M., Steidel, C.\,C., Dickinson, M., Adelberger, K.\,L.,
Giavalisco, M., 1998, ApJ, 508, 539

\bibitem{}
Poggianti B.\,M. \& Barbaro G., 1996, A\&A, 314, 379

\bibitem{}
Poggianti B.\,M. \& Barbaro G., 1997, A\&A, 325, 1025

\bibitem{}
Poggianti, B.\,M., Smail, I., Dressler, A., Couch, W.\,J., Barger, A.\,J.,
Butcher, H., Ellis, R.\,S., Oemler, A., 1999, ApJ, in press (P99)
(astro-ph 9901264)

\bibitem{}
Puget, J.\,L., et al., 1999, A\&A in press (astro-ph 9812039)

\bibitem{}
Sanders, D.\,B., Mirabel, I.\,F., 1996, ARAA, 34, 749

\bibitem{}
Smail, I., Ivison, R., Blain, A., 1997, ApJ, 490, L5

\bibitem{}
Smail, I., Ivison, R., Blain, A., Kneib, J-P, 1998a, ApJ, 507, L21

\bibitem{}
Smail, I., Ivison, R., Blain, A., Kneib, J.-P., 1998b, in \it After the 
Dark Ages: When Galaxies Were Young (the Universe at $2<z<5$), \rm
in press (astro-ph 9810281)

\bibitem{}
Spinrad, H., Dey, A., Stern, D., Dunlop, J., Peacock, J., Jimenez, R., 
Windhorst, R., 1997, ApJ, 484, 581

\bibitem{}
Stasinska, G., 1990, A\&AS, 83, 501

\bibitem{}
Steidel, C.\,C., Adelberger, K.\,L., Dickinson, M., Giavalisco M.,  
Pettini, M., 1998, in {The Birth of Galaxies}, in press (astro-ph 9812167)

\bibitem{}
Steidel, C.\,C., Adelberger, K.\,L., Giavalisco, M., Dickinson, M., 
Pettini, M., 1999, ApJ in press (astro-ph 9811399)

\bibitem{}
Strauss, M.\,A. et al., 1992, ApJS, 83, 29

\bibitem{}
Tran, H.\,D., Brotherton, M.\,S., Stanford, S.\,A., van Breugel, W.,
Dey, A., Stern, D., Antonucci, R., 1998, ApJ in press (astro-ph 9812110)

\bibitem{}
Trentham, N., Kormendy, Sanders, D.\,B., 1999, AnJ, in press  
(astro-ph 9901382)

\bibitem{}
Tresse, L., Maddox, S., Loveday, J., 1999, MNRAS, submitted

\bibitem{}
Treu, T., Stiavelli, M., Walker, A.\,R., et al., 1999, A\&A, in press
(astro-ph 9808282)

\bibitem{}
van Breugel, W., 1999, in {Ultraluminous Galaxies: Monsters and Babies},
in press (astro-ph 9902048)

\bibitem{}
Veilleux, S., Osterbrock, E., 1987, ApJS, 63, 295

\bibitem{}
Veilleux, S., Kim, D.\,C., Sanders, D.\,B., Mazzarella, J.\,M., Soifer, B.\,T.,
1995, ApJS, 98, 171

\bibitem{}
Viallefond, F., Goss, W.\,M., Allen, R.\,J., 1982, A\&A, 115, 373

\bibitem{}
Witt, A.\,N., Thronson, H.\,A., Capuano, J.\,M., 1992, ApJ, 393, 611

\bibitem{}
Wu, H., Zou, Z.\,L., Xia, X.\,Y., Deng, Z.\,G., 1998a, A\&AS, 127, 521 (W98a)

\bibitem{}
Wu, H., Zou, Z.\,L., Xia, X.\,Y., Deng, Z.\,G., 1998b, A\&AS, 132, 181 (W98b)

\bibitem{}
Zabludoff A.\,I., Zaritsky D., Lin H., Tucker D., Hashimoto Y., Shectman S.A.,
Oemler A., Kirshner R.\,P., 1996, ApJ, 466, 104

\bibitem{}
Zaritsky D., Kennicutt R.\,C. Jr., Huchra J.\,P., 1994, ApJ, 420, 87


\end{thebibliography}
\end{document}